\documentclass[10pt,journal,compsoc]{IEEEtran}
\ifCLASSINFOpdf
\else
\fi
\usepackage{graphicx}
\usepackage{multirow}
\usepackage{amssymb}
\usepackage{pifont}
\newcommand{\xmark}{\ding{55}}%
\usepackage{amsmath}
\usepackage{makecell}
\usepackage{rotfloat}
\usepackage{hyperref}
\usepackage{xcolor}
\usepackage{soul}
\usepackage{cite} 

\begin{document}
%
\title{A Procedure and Guidelines for Analyzing Groups of  Software Engineering Replications}
%
%
%
%

\author{Adrian~Santos,
        Sira Vegas,
        Markku~Oivo
        and~Natalia~Juristo
\thanks{A. Santos and M. Oivo are with the M3S (M-Group), ITEE University of Oulu, P.O. Box 3000, 90014, Oulu, Finland, e-mail: \{adrian.santos.parrilla,markku.oivo\}@oulu.fi}
\thanks{S. Vegas and N. Juristo are with the Escuela T\'ecnica Superior de Ingenieros Inform\'aticos, Universidad Polit\'ecnica de Madrid, Campus de Montegancedo s/n, 28660 Boadilla del Monte, Madrid, Spain, e-mail: \{svegas,natalia\}@fi.upm.es}}

%
%

\markboth{February~2019}%
{Shell \MakeLowercase{\textit{et al.}}: Bare Demo of IEEEtran.cls for Computer Society Journals}
%



\IEEEtitleabstractindextext{%
\begin{abstract}

\textbf{Context}: Researchers from different groups and institutions are collaborating on building groups of experiments by means of replication (i.e., conducting groups of replications). Disparate aggregation techniques are being applied to analyze groups of replications. The application of unsuitable techniques to aggregate replication results may undermine the potential of groups of replications to provide in-depth insights from experiment results. \textbf{Objectives}: Provide an analysis procedure with a set of embedded guidelines to aggregate software engineering (SE) replication results. \textbf{Method}: We compare the characteristics of groups of replications for SE and other mature experimental disciplines such as medicine and pharmacology. In view of their differences, the limitations with regard to the joint data analysis of groups of SE replications and the guidelines provided in mature experimental disciplines to analyze groups of replications, we build an analysis procedure with a set of embedded guidelines specifically tailored to the analysis of groups of SE replications. We apply the proposed analysis procedure to a representative group of SE replications to illustrate its use. \textbf{Results}: All the information contained within the raw data should be leveraged during the aggregation of replication results. The analysis procedure that we propose encourages the use of stratified individual participant data and aggregated data in tandem to analyze groups of SE replications. \textbf{Conclusion}: The aggregation techniques used to analyze groups of replications should be justified in research articles. This will increase the reliability and transparency of joint results. The proposed guidelines should ease this endeavor. 
\end{abstract}

}
\maketitle

\IEEEdisplaynontitleabstractindextext

%
\IEEEpeerreviewmaketitle

\section{Introduction}
\label{sec:introduction}

Experiments are commonplace in SE \cite{sjoberg2005survey, stol2018abc, kitchenham2015evidence}. Still, two main shortcomings usually have an impact on their suitability for evaluating the effectiveness of SE technologies \cite{kitchenham2015evidence}: (1) sample sizes are usually small \cite{dybaa2006systematic}, and (2)  results are only generalizable to the configuration of the experimental settings \cite{wohlin2012experimentation}. 

With the aim of increasing the reliability and generalizability of individual experiment results, SE researchers are working on building groups of experiments by means of replication (conducting groups of replications) \cite{munoz2010family, canfora2005family, mouchawrab2011assessing, kosar2012program, abrahao2013assessing}. By collaborating with each other (e.g., sharing experimental material, and assisting each other during the design, execution and/or analysis phase of experiments, etc.), researchers are able to increase the sample size, as well as evaluate the effects of the treatments under different settings. This should increase the reliability of results and their generalizability to different contexts and populations \cite{basili1999building}. 

Groups of replications provide some advantages for evaluating the effectiveness of SE treatments \cite{cooper2009relative, stewart2002ipd, lyman2005strengths, debray2015get, biondi2016umbrella}: (1) access to raw data provides for the use of consistent pre-processing and analysis techniques to analyze each experiment, thereby increasing the reliability of joint conclusions; (2) researchers conducting groups of replications may limit the changes made across the replications in order to increase the internal validity of joint conclusions; (3) joint conclusions are not affected by the detrimental effects of publication bias, as groups of replications do not rely on already published results; (4)  consistent measurement instruments can be used across replications in order to measure participant characteristics with identical methods and scales and, possibly, stratify the results according to such characteristics. 

According to a systematic mapping study (SMS) that we undertook \cite{adrisms}, five techniques are being applied to aggregate groups of SE replications (listed from most to least used): narrative synthesis, aggregated data, mega-trial or stratified  individual participant data, and aggregation of $p$-values. According to the literature of mature experimental disciplines like medicine and pharmacology \cite{brown2014applied, whitehead2002meta}, some aggregation techniques are more suitable than others depending on the characteristics of the group of replications. We observed similar findings when applying the aggregation techniques to analyze a stereotypical group of SE replications \cite{santos2018comparing} (a small group of replications with small and dissimilar sample sizes, opportunistic participant recruitment, different types of subjects, identical experimental designs and response variable operationalizations, and heterogeneous results \cite{adrisms}). The applied aggregation technique definitely had a big impact on the reliability of joint conclusions \cite{santos2018comparing}. 

In view of this, the aim of this study is to answer the following main \textbf{research question}:
\begin{itemize}
    \item{How should groups of SE replications be aggregated?} 
\end{itemize}

To answer this question, we performed a literature review in mature experimental disciplines (medicine and pharmacology) to learn about the techniques applied to aggregate replication results. Along the way, we also noticed some (more profound than expected) differences between groups of replications in SE and medicine. This led to a subsequent literature review of studies with similar circumstances to SE in the fields of medicine, social research, educational research and econometrics. In view of the results, we tailored a \textit{procedure}, with a set of embedded \textit{guidelines}, to facilitate the aggregation of results in groups of SE replications. We apply the proposed procedure to analyze a group of replications in order to illustrate its use. 

This article extends our prior work (i.e., \cite{adrisms, santos2018comparing}) in several ways: (1) by \textit{identifying differences} between the characteristics of groups of replications in the fields of SE and medicine and how such differences may impact the aggregation techniques used to analyze groups of SE replications; (2) by \textit{proposing a step-by-step procedure} to analyze groups of SE replications that takes into account such differences and the typical limitations regarding joint data analysis of groups of SE replications; (3) by \textit{providing a hands-on tutorial} with mathematical formulae and R code snippets to analyze a stereotypical group of SE replications; (4) by \textit{providing a discussion and further pointers} to references indicating how to analyze groups of replications with different experimental designs.

The \textbf{take-away messages} of this research are: 

\begin{itemize}

    \item \textit{Random-effects models should be preferred over fixed-effects models}, especially because many variables may impact SE experiment results, changes are frequent across SE replications, and heterogeneous results are commonplace. Differences between groups of replications in medicine and SE make it inappropriate to directly apply medical guidelines to the analysis of groups of SE replications. In particular, the application of fixed-effects models and traditional statistical thresholds (e.g., the traditional $p$-value of 0.05) in order to detect heterogeneity and moderators does not appear to provide guarantees in SE. 
    
    \item \textit{Avoid narrative synthesis}~\cite{popay2006guidance}, \textit{aggregation of $p$-values}~\cite{borenstein2011introduction} \textit{and mega-trial individual participant data (IPD-MT)}~\cite{field2012discovering}, \textit{and use aggregated data (AD)~\cite{kitchenham2004procedures} and stratified individual participant data (IPD-S)~\cite{simmonds2005meta} in tandem instead}. AD and IPD-S appear to be the most suitable techniques for analyzing groups of SE replications. AD provides intuitive visualizations to convey joint results and straightforward statistics to quantify heterogeneity. IPD-S increases the interpretability of joint results and offers greater statistical flexibility. 
    
       \item \textit{Strive to identify both experiment-level moderators}\footnote{Variables that cause an effect to differ across contexts~\cite{krein2016multi}.} \textit{and participant-level moderators}. AD and IPD-S appear to be good at identifying experiment-level moderators. IPD-S appears to be preferable for identifying participant-level moderators.

        \item \textit{Use the following four-step procedure to analyze a stereotypical group of SE replications}: (1) describe the characteristics of the participants using appropriate descriptive statistics and visualizations; (2) use consistent statistical techniques to pre-process, describe and analyze the data of each replication; (3) select suitable aggregation techniques to provide joint conclusions; and (4) conduct exploratory analyses to identify experiment-level moderators\footnote{Characteristics of the experiments that may be impacting the results, such as the programming language or experimental session length of the experiments.} and participant-level moderators\footnote{Characteristics of the participants that may be impacting the results, such as participant programming or Java experience.}.

\end{itemize}  

The paper has been organized as follows. In Section \ref{background} we provide the background of this study. In Section \ref{research_method} we outline the research method that we followed to elaborate the proposed analysis procedure, and present the stereotypical group of replications that we will use to illustrate its application. In Section \ref{differences} we show the differences between groups of replications in medicine and SE. Then, in Section \ref{limitations} we outline the most common limitations with regard to joint data analysis of groups of SE replications. In Section \ref{procedure} we provide an overview of the four-step analysis procedure that we propose to analyze groups of SE replications. In Sections \ref{step_1}, \ref{step_2}, \ref{step_3} and \ref{step_4} we detail each of the steps of the procedure that we apply to the illustrative group of replications. We outline the threats to validity of this study in Section \ref{threats}, and provide further pointers for the analysis of groups of SE replications with different experimental designs in Section \ref{alternative}. We relate our research to other SE research in Section \ref{related_work}. Section \ref{conclusions} states our conclusions. 

\section{Background}
\label{background}

\subsection{Replication}
\label{replication_background}

The relevance of replication has been widely acknowledged in SE \cite{shull2008role, kitchenham2008role}. Replication has been coupled in SE with the concept of applying a similar experimental procedure to the one applied in a previous baseline experiment on a different sample of participants to generate new raw data \cite{da2014replication, bezerra2015replication}. 

However, two different concepts need to be set apart: replication and reproducibility---or reproduction of results. Researchers who want to reproduce results apply the same analysis procedure followed by the original experimenters to the original raw data with the aim of getting the same results \cite{amann2013software}. Therefore, reproduction has to do with re-analysis of raw data from the baseline experiment \cite{amann2013software}. Replications generate new raw data ---and results--- that can be later combined with the outcomes of other replications to provide joint conclusions. This research focuses on replication.

Different types of replications can be conducted. According to Gomez et al. \cite{gomez2014understanding}, replication types vary along a continuum: from exact replications, following exactly the same experimental configurations as their baseline experiments, to conceptual replications, where the only thing that the replications have in common are the baseline experiment research questions and objectives. Somewhere in between these two extremes lie other replication types, where different elements of the baseline experiment configurations remain unchanged \cite{baldassarre2014replication}.

Laboratory packages were proposed to ease replication across research groups and institutions \cite{shull2004knowledge}. Laboratory packages contain relevant information needed to replicate an experiment \cite{solari2017content}. With a laboratory package, an external group of researchers can reproduce the settings of a baseline experiment, and gather new raw data from a different sample of participants. In addition to sharing laboratory packages, experimenters conducting groups of replications may also collaborate with each other through face-to-face or Internet meetings to plan, design, execute and/or analyze their experiments \cite{juristo2013communication}. This close collaboration may increase the chances of getting similar results across the replications---as experimenter interaction is expected to assure more similar experimental procedures. This should increase the reliability of joint conclusions \cite{borenstein2011introduction}.

Despite even the hardest efforts to conduct exact replications, conflicting results may still pile up \cite{borenstein2011introduction, cumming2013understanding}. It is then that first-hand knowledge of  experiment configurations and  participant characteristics plays a central role. If such information is known, it is easier to hypothesize on the variables that may be behind divergent results. In turn, this is useful for hypothesizing on experiment-level or participant-level moderators that may be influencing the results. It is the above flexibility that leads this research to focus on groups of replications. 

\subsection{Groups of Replications}
\label{characteristics_background}

We conducted a SMS with the aim of learning what aggregation techniques are being used to analyze groups of SE replications \cite{adrisms}. We identified a total of 39 groups of replications that share certain characteristics:

\begin{itemize}

    \item{They are either conducted by \textit{individual researchers} or by \textit{groups of researchers working in close collaboration} across one or multiple research groups, universities and/or institutions. As such, researchers have access to the raw data of all the replications.}

    \item{They are formed \textit{opportunistically}. In other words, a priori plans are not typically set for building groups of replications; each replication comes into being individually without a defined protocol at the inception of the group. As a consequence, replications are aggregated ---generally after having being published individually--- to either increase the reliability of the findings or to elicit moderators (e.g., assessing how the technologies perform for different types of subjects).}

    \item{Most groups of replications are composed of \textit{three to five replications} evaluating the performance of a \textit{binary} treatment (e.g., Method A vs. Method B) on a \textit{continuous outcome of interest} (e.g., productivity measured as LOC per hour). Replications are usually \textit{small}\footnote{Out of consistency with other SE authors \cite{kitchenham2016robust}, small sample size refers throughout this article to experiments involving fewer than 30 subjects.}, have \textit{dissimilar sample sizes}, evaluate the performance of the treatments for \textit{different types of subjects} (e.g., professionals vs. students), have \textit{identical experimental designs and response variable operationalizations}\footnote{Therefore, internal and construct threats to validity are not mitigated or new ones cannot be identified.}, and provide \textit{heterogeneous results}.}
\end{itemize}

\subsection{Aggregation Techniques}
\label{aggregation_background}

Five aggregation techniques have been used in groups of SE replications \cite{adrisms}: narrative synthesis, AD, IPD-MT, IPD-S and aggregation of $p$-values. Thirty-five percent of the groups of replications use more than one aggregation technique. However, they usually serve different purposes: one for providing joint conclusions, and a different one for eliciting moderators. Thus, the groups of SE replications never compare the results achieved with different aggregation techniques for the same objective.

In the following, we review the aggregation techniques used in groups of SE replications starting with the  most, and ending with the least, popular.\footnote{The respective percentage use of the aggregation techniques sum more than 100\% because 16 groups of replications used more than one aggregation technique.} To do this, we rely on the results of the SMS that we undertook \cite{adrisms}.

\textbf{Narrative synthesis} was used to analyze 46\% (18 out of 39) of the groups of replications \cite{adrisms}. In narrative synthesis (also known as semi-quantitative aggregation \cite{borenstein2011introduction}), replication results (in either $p$-value or effect size terms) are combined textually to provide a summary of results. For example, it is common in SE to analyze each replication individually using a $t$-test or a Wilcoxon test and to then provide a textual summary of results of the replications as follows: "...while the results are statistically significant/large/negative in experiments X, Y and Z, they are not in experiment M. This difference of results could have been caused by H, N or K moderator variable...".

\textbf{AD}---commonly known as meta-analysis of effect sizes in SE \cite{kitchenham2004procedures}---was used to analyze 38\% (15 out of 39) of the groups of replications \cite{adrisms}. In AD, all replication effect sizes are first computed from summary statistics, such as means, variances or sample sizes---or from experiment statistical test results---and then combined by means of a meta-analysis model \cite{borenstein2011introduction}. Two different types of meta-analysis models can be fitted: fixed-effects models or random-effects models \cite{borenstein2011introduction}. Fixed-effects models assume that all the experiments estimate a common population effect size and, thus, differences across experiment results arise due to the natural variation of results (i.e., due to the different participant samples in the experiments). On the other hand, random-effects models assume that differences across experiment results arise not just from the natural variation of results, but also from a real heterogeneity of effects. In other words, random-effects models estimate a distribution of population effect sizes rather than a common population effect size.

\textbf{IPD-MT} was used to analyze 33\% (13 out of 39) of the groups of replications \cite{adrisms}. In IPD-MT, the raw data of all the experiments are analyzed jointly as if the raw data came from one big experiment. Since IPD-MT depends on the availability of raw data, researchers typically first analyze each replication individually \cite{field2012discovering} to later perform IPD-MT by pooling and analyzing the raw data of all the replications applying the same statistical test that was used for performing the individual analyses.

\textbf{IPD-S} was used to analyze 15\% (6 out of 39) of the groups of replications \cite{adrisms}. In IPD-S all experiment raw data are analyzed jointly by acknowledging the experiment where the raw data come from. As in AD, two types of IPD-S models can be fitted: fixed-effects models and random-effects models. Commonly used fixed-effects models are linear regression models (e.g., ANOVA) with two factors: Experiment and Treatment \cite{whitehead2002meta}. Commonly used random-effects models are linear mixed models with two factors: Experiment and Treatment \cite{whitehead2002meta}. 

\textbf{Aggregation of $p$-values} was used to analyze 7\% (3 out of 39) of the groups of replications \cite{adrisms}. In aggregation of $p$-values, \textit{one-sided} $p$-values from all replications are pooled together by means of a statistical model such as Fisher's or Stouffer's method \cite{borenstein2011introduction}. Note that $p$-values can be either  available directly (have been previously reported) or computed by researchers (raw data available from each replication is first analyzed to calculate the $p$-values).

\section{Research Method}
\label{research_method}

We began our research by studying the recommendations and guidelines provided in medicine and pharmacology to analyze---and report---groups of replications (i.e., multicenter clinical trials \cite{friedman1998fundamentals}). We resorted to the medical and pharmacological literature because of their longstanding experimental tradition and because SE researchers have previously looked to these disciplines for advice on how to analyze individual experiments \cite{wohlin2012experimentation, juristo2013basics}, how to conduct systematic literature reviews \cite{kitchenham2004procedures} or how to conceptualize new research paradigms, such as evidence-based software engineering \cite{kitchenham2004evidence} and so on. 

Particularly, we began studying the recommendations and guidelines promoted by the Cochrane Association \cite{higgins2008cochrane}, the American Food and Drug Administration \cite{anello2005multicentre}, the guidelines for analyzing multicenter clinical trials (MCTs) provided by the International Conference on Harmonization \cite{lewis1999statistical}, the PRISMA-IPD statement of the EQUATOR Network framework \cite{stewart2015preferred}, and the CONSORT statement for reporting randomized controlled trials \cite{schulz2010consort}. These guidelines for analyzing MCTs are mature and have been widely used. Some have been in use for over 20 years \cite{lewis1999statistical} and others have been referenced thousands of times \cite{schulz2010consort}. 

The above guidelines contain a number of terms (e.g., multicenter, treatment-by-center interaction, etc.), and concepts (e.g., individual participant data, aggregated data, etc.) that we used to drive a subsequent literature review. Throughout the literature review, we came across numerous references to the statistical techniques that can be used to analyze MCTs \cite{debray2015get, fisher2011critical, simmonds2005meta, pincus2011methodological, feaster2011modeling}, meta-analysis \cite{borenstein2011introduction, whitehead2002meta, chen2013applied}, hierarchical linear models \cite{brown2014applied, hox2010multilevel, finch2014multilevel, luke2004multilevel}, and study protocols \cite{de2017rational}. 

After studying the above references, we identified some differences between MCTs and groups of SE replications. According to the meta-analysis and hierarchical linear models literature that we examined, these differences had statistical consequences with respect to results aggregation. This led to another literature review where we discovered studies evaluating data under circumstances more typical of SE: small number of replications with small and unbalanced sample sizes. We found a number of studies from medicine \cite{chu2011comparing, pickering2007analysis}, social research \cite{maas2005sufficient}, educational research \cite{mcneish2016effect} and econometrics \cite{bell2015explaining} studying exactly this. In Section \ref{differences} we outline the differences between groups of SE replications and MCTs, and the statistical consequences of such differences. 

After this second literature review, where we learned about the statistical consequences of the differences between groups of SE replications and MCTs, we revisited the groups of replications that we identified during the SMS reported in \cite{adrisms}. We compiled a list of four common limitations with regard to joint data analysis in groups of SE replications. We outline these limitations in Section \ref{limitations}. 

We developed guidelines to tackle these limitations. We created a four-step analysis procedure with an identical structure to those commonly followed in medicine \cite{higgins2008cochrane, anello2005multicentre,stewart2015preferred, schulz2010consort, lewis1999statistical}. We embedded the guidelines within Steps 3 and 4 of the analysis procedure that we propose (i.e., providing either joint conclusions, or moderator effects, respectively). Before going any further, however, we should clarify that we do not aim to propose an all-encompassing cookbook procedure to analyze all groups of SE replications. Our procedure can be seen as a set of minimum criteria needed to analyze a stereotypical group of SE replications (i.e., with the characteristics outlined in Section \ref{background}). In Section \ref{procedure} we discuss the steps of the proposed analysis procedure, the objectives of each step, the medical guidelines recommending the respective steps, and how we adapted each step to SE by acknowledging differences between MCTs and groups of SE replications, and their common limitations with regard to joint data analysis.

Finally, we outline each of the steps of the proposed procedure. We apply the procedure to analyze a representative group of SE replications to illustrate the procedure.

The chosen group of replications focuses on test-driven development (TDD). One \textit{research question} drives the group of replications: How does TDD affect quality compared to Iterative-Test-Last (ITL)?

The main \textit{independent variable} across all replications is the development approach, with TDD and ITL as treatments. ITL is defined as the reverse-order approach of TDD (following Erdogmus et al. \cite{erdogmus2005effectiveness}). 

All experiments have an identical \textit{experimental design}: an AB repeated-measures design \cite{wohlin2012experimentation} (where subjects first apply ITL, and then TDD at a later date). The \textit{dependent variable} within the group of replications is quality. We measured quality as the percentage of test cases that successfully pass from a battery of test cases that we built for measuring participant solutions. Specifically, we measured quality as follows:
$$ QLTY =\frac{\#Test~Cases(Pass)}{\#Test~Cases(All)}*100\% $$ 

A total of \textit{four replications} were run: three at a multinational online security products company (i.e., F-Secure H, F-Secure K and F-Secure O), and one at UPV, a Spanish university. Six, 11, 7 and 33 subjects participated in each replication, respectively. 

The characteristics of this group of replications are typical in SE: a small number of replications (i.e., four replications, the median number of replications within SE groups) evaluating the performance of a binary treatment (i.e., ITL vs. TDD) on a continuous\footnote{Although the data are measured on a percentage scale (i.e., 0 to 100\%), the approach taken throughout this study is to consider the data as continuous as the total number of test cases is large (i.e., greater than 30 \cite{crawley2012r}). } outcome of interest, with identical response variables and experimental designs, small and dissimilar sample sizes, different types of subjects (i.e., professionals and students) with common knowledge (i.e., week-long training on slicing, unit testing, ITL and TDD) and development culture (test last), using the same environment (i.e., Java, JUnit, Eclipse) and showing heterogeneous results. 

\section{Differences between Groups of SE Replications and MCTs}
\label{differences}

\begin{table*}[t!] \centering 
  \caption{Differences between MCTs and groups of SE replications and statistical consequences.} 
  \label{differences_table} 
\begin{tabular}{p{6cm}p{5.5cm}p{5cm}} \hline \hline 
\textbf{Multicenter Control Trials} & \textbf{groups of SE replications} & \textbf{Statistical consequence} \\ \hline
\checkmark~Identical experimental configurations & \xmark~Opportunistic changes across replications & - Risk of heterogeneity \\ \hline
\checkmark~Rigid \& random participant selection criteria & \xmark~Convenience sampling & - Risk of heterogeneity \\ \hline
\checkmark~Balanced \& adequate sample sizes & \xmark~Unbalanced \& small sample sizes & - Low precision \& power of fixed effects  \\ \hline
\checkmark~Appropriate overall sample size & \xmark~Small overall sample size & - Inability to detect moderators \\ \hline
\end{tabular} 
\end{table*} 

After studying the guidelines and recommendations on analyzing MCTs in medicine and pharmacology, we were skeptical about their direct application for analyzing groups of SE replications due to some relevant differences between MCTs and groups of SE replications. 

For example, MCTs use controlled experiments (participants are  randomly assigned to treatments), while quasi-experiments (where assignment to treatments is non-random) are common in SE. Quasi-experimental designs usually create less compelling support for counterfactual inferences~\cite{william2002experimental}. Quasi-experimental control groups may differ from the treatment condition in many systematic (non-random) ways other than the presence of the treatment. Many of these differences could be alternative explanations for the observed effect and should be ruled out by researchers in order to get a more valid estimate of the treatment effect. By contrast, with random assignment, researchers do not have to think about all these alternative explanations.

Additionally, MCTs tend to have detailed protocols specifying the experimental settings under which all the experiments are to be run and the set of procedures that are to be strictly adhered to during the execution of the experiments \cite{whitehead2002meta, bero1995cochrane, lewis1999statistical}. On the contrary, groups of SE replications are usually created ad hoc \cite{adrisms}. In fact, changes are usually made opportunistically across the replications which are then aggregated to either provide joint results or to investigate moderators. Unfortunately, the changes typically made across SE replications may result in an unexpectedly large variation of results (i.e., statistical heterogeneity of results \cite{borenstein2011introduction}). In practical terms, if statistical heterogeneity materializes, then this is taken as evidence that the treatments may be performing differently across the experiments \cite{borenstein2011introduction}. It may be misleading in this case to apply fixed-effects models to aggregate the results \cite{whitehead2002meta, borenstein2011introduction}---as is typically the case in medicine \cite{whitehead2002meta, anello2005multicentre, phillips2003e9}---. This is because, unlike random-effects methods, fixed-effects models provide a common effect rather than a distribution of effects as a joint conclusion \cite{whitehead2002meta, borenstein2011introduction}. The joint conclusions of fixed-effects model may be especially misleading if results reverse across the experiments \cite{whitehead2002meta, borenstein2011introduction}---as the averaged effect may not, ultimately, be representative for all the experiments. 

We put the absence of protocols in groups of SE replications down to experimental research in SE being less mature than in medicine. Therefore, we regard this as a temporary difference since we expect SE researchers to be convinced by the advantages of developing standardized protocols (e.g., increased internal validity of results \cite{petitti2000meta, friedman1998fundamentals}) and adopt them when conducting groups of replications. 

Besides, stringent and random selection criteria are typically set to recruit the participants in MCT experiments designed to assess the efficacy of new treatments \cite{bero1995cochrane, lewis1999statistical, schulz2010consort}, like specific blood pressure parameters, lack of co-morbid conditions, etc. \cite{bero1995cochrane, lewis1999statistical, schulz2010consort}. This ensures consistent results across sites and helps to minimize the risk of confounding effects impacting results \cite{whitehead2002meta}. Contrariwise, SE replications rarely set stringent selection criteria for recruiting participants. Instead, participants are usually recruited using convenience sampling. Unfortunately, the different characteristics of the participants across the experiments may result in statistical heterogeneity. Once again, this is an obstacle to the application of fixed-effects models for analyzing groups of SE replications. We think that there are two grounds for the absence of strict recruiting criteria in groups of SE replications. First, SE experimental research is less mature and has not yet developed standardized measurement instruments to classify---and include/exclude participants---in SE experiments \cite{falessi2017empirical}. Second, there are differences between the domains of SE and medicine, where SE researchers rarely have the luxury of dismissing participants or an ample array of potential participants. Since we do not expect this to change in the short term, we consider this difference as permanent.

Also, MCTs commonly undertake a planning phase where both participant-level and experiment-level sample sizes are calculated \cite{bero1995cochrane, anello2005multicentre, schulz2010consort}. Participant-level sample sizes define how many subjects are needed, whereas experiment-level sample sizes define how many replications are needed if is only plausible to allocate X subjects to each experiment. This ensures balanced sample sizes across the experiments and proper statistical power for detecting true population effect sizes. On the contrary, sample size estimation phases are rarely undertaken in SE (considering that only one group of SE replications \cite{laitenberger2001internally} provided any sample size requirements calculation \cite{adrisms}). Instead, a small number of replications with small and dissimilar (i.e., convenient) sample sizes are usually run and then aggregated. This sample size estimation phase is feasible within the broader population to which medicine interventions apply as opposed to SE experiments where the population is more restricted. This more contrived sampling frame may prevent groups of SE replications from satisfying statistical power requirements. This places several limitations on the use of fixed-effects models to analyze groups of SE replications. First, fixed-effects models fit many parameters. For example,  parameter estimates may be potentially biased in ANOVA models including an experiment factor, where a different parameter is fitted for each experiment \cite{whitehead2002meta}, due to experiment-level sample size limitations. Second, groups of SE replications tend to have dissimilar sample sizes. This may prevent fixed-effects models achieving the statistical power to detect true treatment effects \cite{chu2011comparing, localio2001adjustments}. We think that the failure in groups of SE replications to pay attention to sample size calculations may be due both to SE experimental research being less mature than medicine and to different participant recruitment opportunities between the SE and medicine domains. Therefore, it is regarded as a permanent difference. 

Finally, the small sample sizes and number of replications in groups of SE replications also impact the detectability of moderators \cite{kraemer2000pitfalls, whitehead2002meta, fisher2011critical}. In particular, larger sample sizes are usually required to detect moderators than to detect treatment effects \cite{kraemer2000pitfalls}. Therefore, it may not be feasible in groups of SE replications to get $p$-values lower than 0.05 in order to claim that there are statistically significant moderator effects. This is especially worrying in the case of experiment-level moderators, as, in most cases, only a few data points are available for moderator detection. It may not be feasible to identify moderators in groups of SE replications unless statistical significance thresholds are adapted (e.g., by increasing them from 0.05 to 0.10 \cite{whitehead2002meta}). Unfortunately, this comes at the cost of a larger proportion of statistical errors \cite{whitehead2002meta, quinn2002experimental}. In our opinion, the inability of groups of SE replications to detect moderators may be due both to SE experimental research being less mature than medicine (as, after all, moderators could be identified if sample size calculations were made \cite{ensor2018simulation}) and differences between SE and medicine (again, in terms of resources). Therefore, it is regarded as a permanent difference. 

Table \ref{differences_table} summarizes the differences between MCTs and groups of SE experiments, and the statistical consequences of such differences for joint data analysis.

\section{Limitations of Groups of SE Replications}
\label{limitations}

We designed an analysis procedure that is identical to the steps followed in medicine and pharmacology to analyze and report MCTs \cite{higgins2008cochrane, anello2005multicentre, lewis1999statistical, stewart2015preferred, schulz2010consort, de2017rational}. We adapted this procedure to groups of SE replications taking into account their typical characteristics in order to overcome common limitations with regard to joint data analysis. After revisiting the groups of SE replications that we identified in our SMS \cite{adrisms}, we came up with a list of four major limitations regarding joint data analysis practices. In the following, they are reviewed one by one.

\textbf{Limitation 1}: Fifty-three percent of the groups of SE replications use either narrative synthesis or aggregation of $p$-values to aggregate replication results \cite{adrisms}. Even though we agree with the use of narrative synthesis and aggregation of $p$-values when the raw data and summary statistics are unavailable or when response variables are incompatible  \cite{popay2006guidance, rodgers2009testing}, we are skeptical about their use when the raw data are available and the replications have identical designs and response variables \cite{adrisms}. In the last analysis, access to the raw data may offer the possibility of providing more informative joint conclusions than just a textual summary of results (narrative synthesis) or a joint $p$-value (aggregation of $p$-values).

\textbf{Limitation 2}: Thirty-three percent of the groups of SE replications were analyzed by means of IPD-MT \cite{adrisms}. This technique may provide misleading results if  participants are more similar within replications than between replications (e.g., when the replications are either run with professionals or with students), or if sample sizes are unbalanced across the treatments and/or replications (e.g., if the replications have different sample sizes and there are missing data). 

\textbf{Limitation 3}: Thirty-eight percent of the groups of SE replications were analyzed by means of AD with standardized effect sizes (such as Cohen's d or the Pearson correlation) \cite{adrisms}. Even though AD with standardized effect sizes can be used to aggregate experiment results in systematic literature reviews (as access to summary statistics or to standardized effect sizes may be guaranteed), we question its use \textit{alone}  when the raw data are available and replications have identical response variable operationalizations\footnote{Groups of SE replications seldom justify the selected aggregation technique \cite{adrisms}. Nevertheless, there appears to be evidence that AD and stratified IPD tend to provide similar results with regard to the provision of joint conclusions \cite{lyman2005strengths, smith2011individual}. The identification of participant-level moderators, however, is a different matter, where stratified IPD comes out on top \cite{fisher2017meta}.}. This is because standardized effect sizes overlook the response variable scales, and, thus, may affect the interpretability of joint conclusions. For instance, how relevant is a joint Cohen's d of 0.3? On the contrary, if the replications have identical response variable operationalizations and access to the raw data is guaranteed---as is typically the case in groups of SE replications \cite{adrisms}---, it may be possible to apply IPD-S \cite{stewart2002ipd, lyman2005strengths, debray2015get} and, thus, interpret results in natural units. This practice has already been applied in SE \cite{runeson2011comparative, krein2016multi, ricca2014assessing} and can lead to more informative joint conclusions. 

\textbf{Limitation 4}: SE researchers rarely acknowledge the limitations of the exploratory analyses that they undertake for identifying moderators. Besides, they usually identify moderators textually (e.g., "as the results are 'statistically significant/positive' in Experiment 1 and not in Experiment 2, this difference between the results could be due to moderator variable X" \cite{adrisms}).

\section{Procedure for Analyzing Groups of Replications}
\label{procedure}

We propose the adoption of a four-step procedure to analyze groups of SE replications. 

\textbf{Step 1. Describe the participants}. We propose to start by describing the participants of the replications. The objectives of this step are not only to \textit{describe the population} to which the results should be generalized, but also to \textit{suggest plausible sources of heterogeneity} that may arise when providing joint conclusions \cite{lewis1999statistical, schulz2010consort}. This step can be further broken down into two main activities:

\begin{itemize}

    \item As typical in MCTs, we propose to start by providing \textit{summary statistics} to describe the characteristics of the participants \cite{lewis1999statistical, schulz2010consort}.

    \item We adapt this step to SE by rounding out the summary statistics with a \textit{profile plot} \cite{alasuutari2008sage} showing the characteristics of the participants averaged across the replications (see Figure \ref{fig:profile_subjects}, Section \ref{step_1}). 
    
\end{itemize}    

\textbf{Step 2. Analyze individual replications}. We propose to pre-process, describe and analyze the data of each replication with consistent statistical techniques. The objectives of this step are \textit{to provide descriptive statistics} to ease the incorporation of results into prospective studies (e.g., by facilitating the recalculation of effect sizes \cite{borenstein2011introduction}), \textit{identify patterns across replication results}, and ensure that \textit{statistical heterogeneity is not introduced by the different methods} used to analyze the replications \cite{silberzahn2018many}. This step can be further broken down into three main activities:

\begin{itemize}

    \item{As in MCTs \cite{lewis1999statistical, stewart2015preferred, schulz2010consort, de2017rational}, we propose to provide \textit{summary statistics and visualizations} (e.g., box plots or violin plots) to describe the data of each replication and use consistent pre-processing steps to remove outliers or replace missing data \cite{schafer2002missing, little2012prevention}.
    
    \item We adapt this step to SE by rounding out the summary statistics and box plots with a \textit{profile plot} \cite{alasuutari2008sage} showing the mean of the treatments across the replications (see Figure \ref{profile}).}

    \item{We analyze each replication with \textit{consistent analyses} (e.g., $t$-test, ANOVA, etc. \cite{lewis1999statistical, schulz2010consort, de2017rational}). This ensures consistency of results across the replications and eases the integration of results in later phases.}    
\end{itemize}

\textbf{Step 3. Aggregate the results}. Following analysis guidelines for MCTs \cite{lewis1999statistical, schulz2010consort, stewart2015preferred}, the results of the individually analyzed replications are aggregated to arrive at \textit{joint conclusions}. The objective of this step is to \textit{increase the reliability of joint conclusions}. We adapt this step to SE by proposing three guidelines, each specifically tailored to address Limitations 1-3 discussed in Section \ref{limitations}:  

\begin{itemize}

    \item Guideline 1 draws upon arguments from groups of data analysis experts in mature experimental disciplines \cite{bero1995cochrane, anello2005multicentre, lewis1999statistical, stewart2015preferred} and the latest recommendations provided by statistical reformers and associations \cite{wasserstein2016asa, cumming2013understanding, mcelreath2015statistical} to suggest avoiding the use of narrative synthesis and aggregation of p-values to provide joint conclusions. 
    
    \item Guideline 2 recommends avoiding IPD-MT by default. Identical advice has been already provided in mature experimental disciplines such as medicine and pharmacology \cite{kraemer2000pitfalls, abo2013individual, feaster2011modeling, stewart2015preferred, kahan2013analysis}. 
    
    \item Guideline 3 draws on arguments from various resources regarding linear mixed models \cite{brown2014applied, hox2010multilevel}, references comparing the performance of IPD-S and AD \cite{stewart2002ipd}, and articles comparing the performance of various IPD-S models \cite{burke2017meta} to encourage the use of both AD and IPD-S \textit{in tandem} to analyze groups of SE replications. A similar recommendation has already been provided in mature experimental disciplines such as medicine and pharmacology \cite{tierney2015individual, smith2011individual}. 
    
    \end{itemize}

We also adapt this step to SE by proposing the use of random-effects models (rather than the fixed-effects models typically used in MCTs \cite{whitehead2002meta, anello2005multicentre}) in the two activities into which this step is divided: \textit{apply AD} and \textit{apply IPD-S}. Similar advice to this has also been given under similar circumstances in other disciplines such as the social sciences \cite{borenstein2011introduction, greco2013meta, clark2015should}.

\textbf{Step 4. Conduct exploratory analyses}. As in MCTs \cite{schulz2010consort, stewart2015preferred, lau1998summing, tierney2015individual, de2017rational}, exploratory analyses should be conducted after providing joint conclusions. The objective of this step is to \textit{identify experiment-level and participant-level moderators} that may be behind the statistical heterogeneity commonly present in groups of SE replications. We adapt this step to SE by developing three new guidelines to address Limitation 4 discussed in Section \ref{limitations}. To do this, we rely on the recommendations provided in references on data analysis in the social sciences, biology, and medicine:

\begin{itemize}

    \item Guideline 4 provides guidance on how to identify experiment-level moderators by means of AD and IPD-S \cite{lau1998summing, quinn2002experimental, abo2013individual, fisher2011critical, cooper2009relative, higgins2001meta}.

    \item Guideline 5 provides guidance on how to identify participant-level moderators by means of IPD-S \cite{lau1998summing, quinn2002experimental, abo2013individual, fisher2011critical, fisher2017meta}.

    \item Guideline 6 outlines the limitations of exploratory analyses \cite{schulz2010consort, stewart2015preferred, lau1998summing, tierney2015individual, quinn2002experimental, whitehead2002meta, cumming2013understanding}.

\end{itemize}

We also adapt the procedure for identifying moderators to SE by suggesting an increase in the statistical significance threshold from 0.05 to 0.10---at the greater risk of a larger proportion of statistical errors \cite{whitehead2002meta}. We also suggest that less attention be paid to $p$-values, evaluating instead the relevance of moderator effect sizes---and their respective 95\% CIs. These recommendations account for the typically low number of replications and small sample sizes of groups of SE replications, which are an obstacle to moderator detection \cite{adrisms}. The latter two adaptations are used again in the first two activities of this step: \textit{identify experiment-level moderators} and \textit{identify participant-level moderators} and acknowledged in the last activity of this step, \textit{acknowledge limitations of exploratory analyses}.

Table \ref{steps_analysis} summarizes the steps of the procedure that we propose for analyzing groups of SE replications, the objectives of each step, and their associated activities. Table \ref{traceability} links each of the proposed steps with the guidelines from medicine recommending the respective step, and the adaptation that we made for SE. 

\begin{table*}[t!] \centering 
  \caption{Procedure for analyzing groups of SE replications and objectives.} 
  \label{steps_analysis} 
\begin{tabular}{lll} \hline \hline 
\textbf{Step} & \textbf{Objectives} & \textbf{Activity} \\ \hline 
\multirow{2}{3cm}{1. Describe participants} & - Inform about the population under assessment & 1.1. Provide summary statistics \\ 
& - Hypothesize on possible sources of heterogeneity & 1.2. Provide profile plot \\ \hline
\multirow{3}{3cm}{2. Analyze individual replications} & - Ease incorporation of results into prospective studies & 2.1. Provide summary statistics and visualizations \\ 
& - Identify patterns across replication results & 2.2. Provide profile plot\\
& - Avoid heterogeneity due to different analysis procedures & 2.3. Perform consistent individual analyses\\ \hline
\multirow{2}{3cm}{3. Aggregate results} & \multirow{2}{*}{- Maximize informativeness of joint conclusions} & 3.1. Apply AD \\
& & 3.2. Apply IPD-S \\ \hline
\multirow{3}{3cm}{4. Conduct exploratory analyses} & \multirow{2}{*}{- Identify experiment-level moderators} & 4.1. Identify experiment-level moderators \\
& \multirow{2}{*}{- Identify participant-level moderators} & 4.2. Identify participant-level moderators \\ 
& & 4.3. Acknowledge limitations of exploratory analyses \\ \hline \\
\end{tabular} 
\end{table*} 

\begin{table*}[t!] \centering 
  \caption{Mapping of proposed procedure steps to references in medicine and adaptations made for SE.} 
  \label{traceability} 
\begin{tabular}{llll} \hline \hline 
\textbf{Step} & \textbf{Recommended in...}& \textbf{Adaptation to SE} & \textbf{Adapted from...}\\ \hline 

\multirow{1}{3cm}{1. Describe participants} & \cite{lewis1999statistical, schulz2010consort} & \checkmark Provide profile plot & \cite{alasuutari2008sage} \\ \hline

\multirow{2}{3cm}{2. Analyze individual replications} & \multirow{2}{2cm}{\cite{lewis1999statistical, stewart2015preferred, schulz2010consort, de2017rational}} &  \multirow{2}{*}{\checkmark Provide profile plot} & \multirow{2}{*}{\cite{alasuutari2008sage}}\\ 
& & \\ \hline

\multirow{4}{3cm}{3. Aggregate results} & \multirow{4}{2cm}{\cite{anello2005multicentre, lewis1999statistical, stewart2015preferred, schulz2010consort, feaster2011modeling}} &  \checkmark Avoid narrative synthesis \& aggregation of p-values & \cite{cumming2013understanding, bero1995cochrane, wasserstein2016asa, mcelreath2015statistical} \\ 
& & \checkmark Avoid IPD-MT & \cite{kraemer2000pitfalls, abo2013individual, kahan2013analysis} \\
& & \checkmark Use AD \& IPD-S in tandem  & \cite{stewart2002ipd, brown2014applied, hox2010multilevel, burke2017meta, tierney2015individual, smith2011individual}\\
& & \checkmark Use random-effects models & \cite{whitehead2002meta} \\ \hline

\multirow{5}{3cm}{4. Conduct exploratory analyses} & \multirow{5}{2cm}{\cite{schulz2010consort, stewart2015preferred, lau1998summing, tierney2015individual, de2017rational, fisher2011critical}} & \checkmark Use AD \& IPD-S to assess experiment-level moderators & \cite{cooper2009relative, quinn2002experimental, abo2013individual, higgins2001meta}\\
& & \checkmark Use IPD-S to identify participant-level moderators & \cite{quinn2002experimental, abo2013individual, fisher2017meta} \\
& & \checkmark Acknowledge limitations of exploratory analyses & \cite{whitehead2002meta, cumming2013understanding, quinn2002experimental} \\
& & \checkmark Increase statistical threshold & \cite{adrisms, whitehead2002meta} \\
& & \checkmark Evaluate effect size and 95\% CI & \cite{adrisms}\\
\hline
\end{tabular} 
\end{table*}

In the next four sections we outline the four steps of the procedure that we propose for analyzing groups of SE replications. For illustrative purposes, we apply the respective step to analyze the stereotypical group of replications described in Section \ref{research_method}. As Step 1 (i.e., describe the participants) and Step 2 (i.e., analyze individual replications) require no further explanation, we merely apply the steps to analyze the illustrative group of replications in Sections \ref{step_1} and \ref{step_2}. As Step 3 (i.e., aggregate the results) and Step 4 (i.e., conduct exploratory analysis) embed a set of guidelines that require further explanation, we first develop the guidelines and then go on to illustrate their application to the group of replications in Sections \ref{step_3} and \ref{step_4}.

\section{Step 1: Describe the participants}
\label{step_1}

\textbf{Activity 1.1. Provide summary statistics}. The provision of summary statistics of the characteristics of the participants offers information about the population to which the results are to be generalized.

\textbf{Example}. Table \ref{median_experiences} shows the means and standard deviations of participant programming, Java, unit testing and JUnit  experience (measured by self-assessment as inexperienced, novice, intermediate and expert) across the replications. We do not find any clear patterns for averaged participant-level characteristics (e.g., averaged experience levels for F-Secure O and F-Secure H participants alternate between programming and Java) across replications.

\begin{table}[h!] \centering 
  \caption{Descriptive statistics for participant characteristics.} 
  \label{median_experiences} 
\begin{tabular}{lcccc} \hline \hline  
\textbf{Experiment} & \textbf{Prog.} & \textbf{Java} & \textbf{Unit} & \textbf{JUnit} \\ \hline 
F-Secure H&3.67 (0.52) & 2.33 (1.21) & 2.17 (0.98) & 2.17 (1.17) \\ 
F-Secure K& 2.91 (0.70) & 1.82 (0.87) & 1.64 (0.5) & 1.27 (0.47) \\ 
F-Secure O& 3.29 (0.76) & 2.71 (1.11) & 2.71 (0.76) & 2 (0.82) \\ 
UPV &2.36 (0.57) & 1.88 (0.60) & 1.04 (0.20) & 1 (0) \\ 
\hline 
\end{tabular} 
\end{table}

\textbf{Activity 2.1. Provide profile plot}. The provision of a profile plot showing the mean experience of the participants across replications may enhance the understandability of the summary statistics, help to convey the variability of participant characteristics across replications, ease the identification of patterns in the characteristics of the participants across experiments, and help to identify potential sources of heterogeneity.

\begin{figure}[h!]
    \centering
    \includegraphics[width=9cm,keepaspectratio]{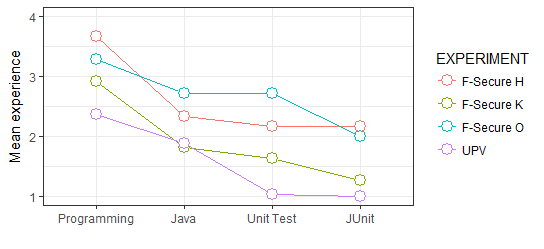}
    \caption{Profile plot for subject experience.}
    \label{fig:profile_subjects}    
\end{figure}

\textbf{Example}. Figure \ref{fig:profile_subjects} shows the profile plot of the illustrative group of replications. Figure \ref{fig:profile_subjects} indicates that there is an observable decreasing trend in averaged participant experience across replications: participants have relatively more experience with programming and Java than with unit testing or JUnit across all the replications. There appears to be a noticeable heterogeneity of averaged participant experience. This may result in statistical heterogeneity when providing joint conclusions.

\textbf{Summary of example.} A heterogeneous group of developers participates within the group of replications, with the most senior developers at F-Secure H and F-Secure O, and the junior at UPV.

\section{Step 2: Analyze individual replications}
\label{step_2}

\textbf{Activity 2.1. Provide summary statistics and visualizations.} The summary statistics and box plots provide information about the distribution of the data, facilitate the incorporation of results into prospective studies, and minimize the heterogeneity of results due to the application of different analysis techniques. 

\begin{figure}[h!]
    \centering
    \includegraphics[width=9cm,keepaspectratio]{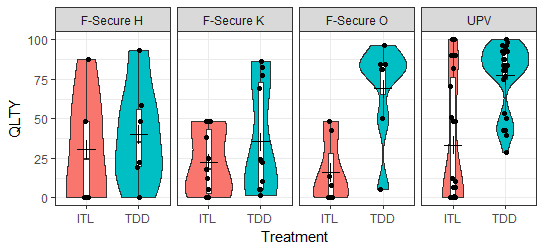}
    \caption{Box plot and violin plot: ITL vs. TDD.}
    \label{violin}    
\end{figure}

\textbf{Example.} Table \ref{descriptive} shows the descriptive statistics for QLTY with ITL and TDD across all replications. The respective box plots---and violin plots---are shown in Figure \ref{violin}.

\begin{table}[h!] \centering 
  \caption{Descriptive statistics for QLTY: ITL vs. TDD.} 
  \label{descriptive} 
\begin{tabular}{llccccc} \hline \hline  
\textbf{Experiment} & \textbf{Treat.} & \textbf{N} & \textbf{Mean} & \textbf{Corr} & \textbf{SD} & \textbf{Median} \\ \hline 
\multirow{2}{*}{F-Secure H} & ITL & $6$ & $30.71$ & \multirow{2}{*}{0.59} & $36.58$ & $24.16$ \\ 
& TDD & $6$ & $40.23$ & &$33.43$ & $35.34$ \\ \hline
\multirow{2}{*}{F-Secure K} & ITL & $11$ & $22.17$ & \multirow{2}{*}{0.42} & $20.44$ & $17.98$ \\ 
& TDD & $11$ & $35.42$ & & $35.40$ & $22.41$ \\  \hline
\multirow{2}{*}{F-Secure O} & ITL & $7$ & $16.05$ & \multirow{2}{*}{0.52} & $20.81$ & $7.87$ \\ 
& TDD & $7$ & $68.97$ & &$31.53$ & $81.03$ \\  \hline
\multirow{2}{*}{UPV} & ITL & $31$ & $33.38$ & \multirow{2}{*}{0.47} &$39.79$ & $6.74$ \\ 
& TDD & $29$ & $77.16$ & & $21.04$ & $83.93$ \\ \hline
\end{tabular} 
\end{table} 

As Figure \ref{violin} shows, TDD appears to outperform ITL in all replications. However, while the difference in performance between ITL and TDD is small for F-Secure H and F-Secure K, the difference appears to be larger for F-Secure O and UPV. Noticeable---and similar---correlations appear to have materialized among the QLTY scores of the participants across replications (i.e., correlations around 0.5). This will result in greater statistical power (i.e., smaller effect size variances) when analyzing each replication and calculating their respective effect sizes \cite{cumming2013understanding, morris2002combining}. Finally, some data points (at the bottom of the distributions) for F-Secure O or UPV may be considered outliers. However, due to the already small sample sizes of the replications and missing data for UPV (two participants have data for TDD only, and another two have none), we do not remove any potential outlier from the data analysis.

\textbf{Activity 2.2. Provide profile plot.} A profile plot to complement the descriptive statistics provides a bird's eye-view of the data and helps identify patterns in results. 

\begin{figure}[h!]
    \centering
    \includegraphics[width=9cm,keepaspectratio]{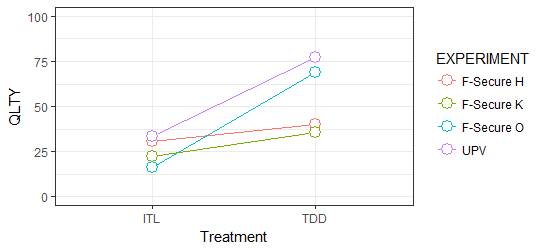}
    \caption{Profile plot: ITL vs. TDD.}
    \label{profile}    
\end{figure}

\textbf{Example.} A profile plot showing the mean QLTY score per treatment across replications is provided in Figure \ref{profile}.

As Figure \ref{profile} shows, TDD appears to outperform ITL across all replications as observed in the violin plot above. The extent to which TDD outperforms ITL varies widely across replications (see the different slopes of the lines). The different slopes indicate that \textit{there may be heterogeneity in the group of replications}. Besides, there is no apparent pattern between ITL and TDD mean QLTY scores: the larger improvements with TDD over ITL (see the lines with the highest slopes) are achieved in the replications with the lowest and highest ITL mean scores (i.e., UPV and F-Secure O replications, respectively). By chance, they are the replications with the most novice and senior developers, respectively. Thus, \textit{we cannot hypothesize, in principle, on any moderator that may be impacting results}.

\textbf{Activity 2.3. Perform consistent individual analyses.} An analysis of the replications with consistent statistical methods ensures that differences across experiment results are not due to the use of different analysis procedures. 

\textbf{Example.} Since all replications have an identical AB repeated-measures experimental design \cite{wohlin2012experimentation}, we analyze each of them with a dependent $t$-test \cite{field2013discovering}. Table \ref{results_individual} shows the results of the individual $t$-tests performed.

\begin{table}[h!] 
\begin{center}
  \caption{Individual analyses: ITL vs TDD.}
  \label{results_individual} 
\begin{tabular}{lccc} 
\hline \hline
\textbf{Experiment} & \textbf{Estimate} & \textbf{95\% CI} & \textbf{$p$-value}   \\  \hline 
F-Secure H & 9.52 & (-19.58, 38.62) & 0.483 \\  
F-Secure K & 13.26 & (-7.26, 33.77) & 0.193 \\ 
F-Secure O & 52.91 & (30.44, 75.39) & \textbf{$<$0.001} \\   
UPV & 42.31 & (29.02, 55.62) & \textbf{$<$0.001}\\ \hline 
\end{tabular} 
\end{center}
\end{table} 

As Table \ref{results_individual} shows, TDD outperforms ITL in all replications. Besides, the difference in performance between TDD and ITL is large and statistically significant for F-Secure O and UPV but not for F-Secure H and F-Secure K. 

\textbf{Summary of example.} TDD outperforms ITL in all replications. However, the extent to which TDD outperforms ITL seems largely dependent upon site. 

\section{Step 3: Aggregate the results}
\label{step_3}

In Sections \ref{guideline_one}, \ref{guideline_two} and \ref{guideline_three}, we outline the three guidelines that we propose to overcome the most common limitations of groups of SE replications when providing joint conclusions. The description includes its application to the illustrative group of replications.

\subsection{Avoid narrative synthesis and aggregation of $p$-values}
\label{guideline_one}

\subsubsection{Perils of narrative synthesis and aggregation of $p$-values}

Although $p$-values are commonly used  to evaluate the statistical significance of results, numerous criticisms have been made with respect to their inappropriate use across the sciences \cite{nickerson2000null, cohen1994earth}. The dichotomization of evidence possibly arising as a result of the indiscriminate use of statistical thresholds (such as 0.05 \cite{cohen1994earth}), and the inability of $p$-values to convey the relevance of results (because they confound sample size and effect size \cite{nickerson2000null}) are just two well-known criticisms of $p$-values \cite{nickerson2000null}. As effect sizes and 95\% CIs can also be used to assess the statistical significance of results (i.e., if the 95\% CI of the effect size does not cross 0, then results are statistically significant), some authors have suggested that effect sizes and 95\% CIs should be used instead of $p$-values \cite{wasserstein2016asa, cumming2013understanding, mcelreath2015statistical}.

Bearing this in mind, neither narrative synthesis nor aggregation of $p$-values appear to be suitable for providing joint conclusions in groups of SE replications: narrative synthesis yields neither an effect size nor a $p$-value (it merely provides a textual summary of results), whereas aggregation of $p$-values provides a joint $p$-value but not an effect size.

Narrative synthesis and aggregation of $p$-values have another shortcoming: narrative synthesis weights each replication subjectively, while aggregation of $p$-values weights each replication within the joint conclusion identically \cite{borenstein2011introduction}. Both types of weighting may be undesirable in groups of replications with different sample sizes. For example, larger (in principle, more precise) replications may have a greater weight than small replications within the joint conclusion, industrial (in principle, more representative) replications may have a greater weight than academic replications, and higher quality experiments may have a greater weight within the joint conclusion, etc. 

Finally, narrative synthesis has another relevant shortcoming when providing joint conclusions. Very often non-significant results lead to a joint statistically significant result \cite{borenstein2011introduction}. However, the joint conclusion of narrative synthesis would be non-significant if there are more non-significant results than significant results (as non-significant is the winner \cite{borenstein2011introduction}). Narrative synthesis has been known since the 1980s to have low statistical power \cite{hedges1980vote, borenstein2011introduction}.

\subsubsection{Application to the illustrative group of replications}

\textbf{Narrative synthesis} is simply applied by providing a textual summary of results of the replications (i.e., their effect sizes and $p$-values \cite{borenstein2011introduction}).

To apply narrative synthesis to the illustrative group of replications, the procedure is as follows: "...even though TDD outperformed ITL in all the replications, the extent of such outperformance was largely dependent upon site. Besides, the difference in performance between TDD and ITL was statistically significant only for F-Secure O and UPV. Thus, conflicting results materialized in terms of statistical significance: two replications provided non-significant results, while two others provided significant results. As an identical number of replications point in opposite directions---i.e., non-significant vs. significant, \textit{no final claims} can be made about the statistical significance of results. More replications are needed to argue the statistical significance, and practical relevance of results." 

\textbf{Aggregation of $p$-values} procedures typically involve \cite{borenstein2011introduction, whitehead2002meta}: (1) the individual analysis of each replication with a one-sided statistical test; and (2) the later combination of the resulting $p$-values by a statistical technique like Fisher's method. 

We first analyzed each replication independently by means of a \textit{one-sided} dependent $t$-test \cite{field2013discovering}. Then, we used Fisher's method \cite{borenstein2011introduction} to pool together the $p$-values of all the replications. The result is a statistically significant difference between TDD and ITL as a joint conclusion ($\tilde{\chi}^2$=47.13; $df$=8; $p<$0.001). Thus, the difference in performance between TDD and ITL \textit{is statistically significant} \textit{in at least one replication}. However, this was already known before aggregating the results (as F-Secure O and UPV's results were already statistically significant). 

\textbf{Summary of example.} Neither narrative synthesis nor aggregation of $p$-values provide informative joint conclusions. Narrative synthesis fails to provide a joint effect size or $p$-value and is not able to provide final claims since there are two significant results versus two non-significant results in our example. Aggregation of $p$-values fails because it provides a joint conclusion that was already known before results aggregation. 

\subsubsection{Guideline 1: Avoid narrative synthesis and aggregation of p-values}

Avoid narrative synthesis and aggregation of $p$-values to provide joint conclusions. 

What impact may this guideline have on the findings of joint analyses of groups of SE replications? More informative joint conclusions could have been obtained for 53\% of the groups of replications (i.e., groups that applied either narrative synthesis or aggregation of $p$-values, see Section \ref{limitations}). Not applying weak aggregation techniques should enhance the findings of groups of SE replications.

\subsection{Avoid IPD-MT}
\label{guideline_two}

\subsubsection{Perils of IPD-MT}
 
IPD-MT should be avoided on two grounds. First, it may be \textit{underpowered} compared to an identical IPD-S model including a factor accounting for the experiment \cite{chu2011comparing, kahan2013assessing}. In other words, IPD-MT may provide a statistically non-significant result when it should be statistically significant. Second, IPD-MT may provide \textit{biased results} \cite{kraemer2000pitfalls, kwok2008analyzing} when data are unbalanced across treatments and replications (which may be the case in groups of replications with missing data and different sample sizes) and subjects are more similar within, than between, replications (which may be the case when either professionals or students participate in the replications). Here we illustrate the perils of IPD-MT with an intuitive extreme example where it provides a biased result. Like Kraemer's example to illustrate the perils of IPD-MT \cite{kraemer2000pitfalls}, we produce our example by means of simulation \cite{cumming2013understanding}. 

Particularly, let us simulate two hypothetical replications comparing the performance of two technologies (e.g., Technology A vs. Technology B) on a continuous outcome of interest (e.g., quality). For simplicity's sake, let us suppose that the replications have an identical experimental design: an AB between-subjects design (i.e., a design where each participant is assigned to either Technology A or B). It is straightforward to simulate a group of replications with such characteristics using random draws from the data distributions that simulate the quality scores achieved with Technologies A and B across the replications \cite{cumming2013understanding}. Each random draw will represent the quality score achieved by a hypothetical (i.e., simulated) participant. SE data may follow a myriad of data distributions \cite{kitchenham2016robust}. For illustrative purposes, we simulate the performance of Technologies A and B with normal distributions \cite{cumming2013understanding}, although many other data distributions could have been used and have obtained the same results \cite{cumming2013understanding}. Table \ref{tab:unbalanced_settings_two} shows the normal distributions that we use to simulate the performance of Technologies A and B across the replications, and the sample sizes of each of the groups (i.e., the number of participants assigned to either Technology A or B) across the replications.

\begin{table}[h]
\begin{center}
\caption{Example: two simulated replications.}
\label{tab:unbalanced_settings_two}
\begin{tabular}{ l | l | l | l} \hline \hline
 & & \textbf{Technology A} & \textbf{Technology B}  \\ \hline
 \multirow{2}{*}{\textbf{Exp. 1}} & QLTY & $\mathcal{N}(20,10^2)$ & $\mathcal{N}(30,10^2)$\\ \cline{2-4}
 & Sample Size  &  90 & 10  \\ \hline 
  \multirow{2}{*}{\textbf{Exp. 2}} & QLTY & $\mathcal{N}(60,10^2)$ & $\mathcal{N}(70,10^2)$\\ \cline{2-4}
 & Sample Size  &  10 & 90  \\ \hline 
\end{tabular}
\end{center}
\end{table}

As Table \ref{tab:unbalanced_settings_two} shows, we aim to simulate two highly unbalanced replications (i.e., 90 subjects assigned to Technology A, and 10 to Technology B in Experiment 1, and vice versa in Experiment 2). Additionally, we aim to simulate a circumstance where the mean difference in performance between Technologies B and A is expected to be around 10 in both replications (i.e.,  30-20 in Experiment 1 and 70-60 in Experiment 2), and the participants are more similar within, than between, replications (as they achieve either much larger or much smaller scores with either Technology A or B depending upon the replication in which they participate). These are the exact circumstances under which IPD-MT provides biased results.

As the difference in performance between Technologies B and A in both replications is around 10, we would expect the difference in performance to be similar for joint results.

We analyzed the data with both IPD-S and IPD-MT (i.e., ANOVA models that did not did not include Experiment as a factor, respectively). IPD-S provides an estimate close to the expected ($M=9.55$). IPD-MT provides an estimate that deviates from the expected ($M=41.17$). This is because IPD-MT does not take into account the experiment that is the source of the data and, instead, assumes that all the data come from a single "big" experiment. As such, IPD-MT is unaware that most subjects contributing towards the mean quality score with Technology A (90/100) come from Experiment 1 (with mean scores of 20), whereas most subjects contributing towards Technology B (90/100) come from Experiment 2 (with mean scores of 70). As a result, the unbalance of subjects across the treatments and the dissimilarities of participant scores across the replications distort IPD-MT results (i.e., by providing a much larger difference of results than expected). The larger the unbalance across treatments and/or sample sizes across the replications, the more biased IPD-MT results will be.

\subsubsection{Application to the illustrative group of replications}

To \textbf{apply IPD-MT}, the raw data of all replications are pooled together as if they come from one big experiment and then the same statistical model as used to analyze each experiment individually is applied \cite{kraemer2000pitfalls, abo2013individual, feaster2011modeling}.

We apply a dependent $t$-test to analyze the raw data of all the replications together. IPD-MT provides a joint estimate equal to $M=33.67$ and a $p$-value$<$0.001. Thus, the difference in performance between TDD and ITL is large---insofar as QLTY ranges from 0 to 100---and statistically significant.

To \textbf{apply IPD-S}, raw data from all replications are pooled together, and then two factors ---Treatment and Experiment--- are used \cite{kraemer2000pitfalls, abo2013individual, feaster2011modeling}. In other words, a linear regression model (e.g., ANOVA) that takes into account the source of the raw data is fitted.

We applied an ANOVA with Treatment and Experiment as factors to analyze the raw data of the illustrative group of replications. IPD-S provides a joint estimate equal to $M=34$ and a $p$-value$<$0.001. Thus, the difference in performance between TDD and ITL is large---insofar as QLTY ranges from 0 to 100---and statistically significant.

So, both IPD-MT and IPD-S provide similar results in the illustrative group of replications (i.e., a large and statistically significant result). This is because data are perfectly balanced within replications (as the replications are AB repeated-measures designs), missing data have a relatively low impact on results (as just a few subjects have missing data for UPV), and subject scores are similar across the replications---note that the mean scores for ILT are clustered around 25 (see Figure \ref{profile}, Section \ref{step_2}). However, this cannot be guaranteed in all circumstances.

\subsubsection{Guideline 2: Avoid IPD-MT}

Avoid IPD-MT due to its potential to provide biased or underpowered results. 

What impact may this guideline have on the findings of joint analyses of groups of SE replications? Thirty-three percent of the groups of replications (i.e., where IPD-MT was applied to provide joint conclusions, see Section \ref{limitations}) could have arrived at less biased and less underpowered joint conclusions. Not applying IPD-MT should enhance the findings of groups of SE replications.

\subsection{Use AD and IPD-S in tandem for joint analysis}
\label{guideline_three}

\subsubsection{Benefits of using AD plus IPD-S}
\label{benefits_tandem}

AD and IPD-S are complementary in some respects. While AD provides certain advantages for analyzing groups of SE replications, IPD-S provides others. \textbf{Some of the advantages of AD over IPD-S are}:

\begin{itemize}

	\item{AD can be used to analyze groups of replications with \textit{different response variables} (e.g., by computing standardized effect sizes such as Cohen's d \cite{borenstein2011introduction}). On the contrary, IPD-S can only be applied whenever identical response variable scales are used \cite{whitehead2002meta, stewart2002ipd}. Thus, if response variables change across the replications, AD may be the only available option.}

    \item{AD provides \textit{intuitive visual summaries of results} (i.e., forest plots) that have been commonly used in SE to synthesize the findings of experiments gathered by means of systematic literature reviews \cite{kitchenham2004procedures}. On the contrary, less standardized visualizations are available for IPD-S (e.g., 95\% CI plots, error bars, etc. \cite{cumming2013understanding}). Thus, the appeal and familiarity of forest plots in SE is a plus for AD over IPD-S.}

	\item{AD is useful for \textit{interpreting the heterogeneity of results with straightforward statistics and tests} (e.g., the $I^2$ statistic and the $Q$-test \cite{borenstein2011introduction}). Besides, rules of thumb are also available for interpreting the $I^2$ statistic (i.e., 25\%, 50\% and 75\% for small, medium, and large heterogeneity, respectively \cite{borenstein2011introduction}). On the contrary, IPD-S may require either: (1) the standard deviation of results to be contrasted against the joint result; or (2) fixed-effects models (such as ANOVA) to be used with Treatment by Experiment interaction terms to be able to claim that results are heterogeneous when the interaction is statistically significant \cite{whitehead2002meta}. However, we do not encourage the latter procedure in groups of SE replications, which are typically small. Therefore, this method of heterogeneity detection would be underpowered \cite{whitehead2002meta}.}

\end{itemize}

On the other hand, \textbf{IPD-S has some advantages over AD}:
\begin{itemize}
    \item{IPD-S can simultaneously assess \textit{the difference in performance} between the treatment and the control group (like AD), as well as the \textit{performance of the control group} in order to weight their \textit{relative size in natural units}. For example, if the difference in performance between the means of the treatment group and the control group is equal to 20, and the mean performance of the control group is equal to 20, then the treatment \textit{doubles} the performance of the control. On the contrary, AD commonly relies on standardized effect sizes (e.g., Cohen's d \cite{borenstein2011introduction}) to convey the difference in performance between the treatment and control. This may affect the interpretability of results: how \textit{relevant} is a Cohen's d of 0.3? }
    \item{IPD-S can simultaneously assess \textit{the effect of multiple factors and their interactions on results} (e.g., the effects of the treatments, the tasks, and their interaction in ANOVA models \cite{wohlin2012experimentation}). On the contrary, AD is commonly used to perform \textit{pairwise comparisons} between treatments (e.g., Treatment A vs. Treatment B \cite{borenstein2011introduction}). Thus, IPD-S is more flexible than AD for analyzing groups of replications when multiple factors are of interest or the results depend upon interaction terms (e.g., when the effects of the treatment reverse depending upon the task being developed).}
    \item{Some IPD-S models such as LMMs can be used to \textit{analyze groups of replications with missing data}---provided that the data can be assumed as missing at random \cite{brown2014applied}. On the contrary, the calculation of effect sizes---and their respective variances---using AD rests on the assumption of complete observations (otherwise, it would not be possible to compute the variances of some effect sizes for repeated-measures designs \cite{borenstein2011introduction}). If there are missing data, researchers performing AD to calculate effect sizes and their respective variances may have to either exclude participants with missing data or rely on advanced imputation techniques for their inclusion (see Section \ref{alternative}). Thus, if there are drop-outs or protocol deviators across the replications, IPD-S models such as LMMs may come in handy \cite{twisk2013multiple}.}
\end{itemize}

Therefore, the application of both techniques in tandem takes advantage of the strengths of each one. Regarding the type of model to be used, both AD and IPD-S are statistical procedures that deliver a \textit{weighted average} of experiment results as a joint conclusion \cite{whitehead2002meta, borenstein2011introduction}. The \textit{weight}---or contribution---of each experiment towards the joint conclusion is proportional to either \textit{the sample size of the experiment}---if a fixed-effects model is used---or to \textit{the sample size of the experiment} and the \textit{statistical heterogeneity of results} (i.e., the variation of results that cannot be explained by natural variation)---if a random-effects model is used \cite{borenstein2011introduction, brown2014applied}.\footnote{Assuming a common variance across all replications.} Besides, if results are more heterogeneous, the weights of all the experiments within the joint conclusion will be more alike. Intuitively, as each experiment may be estimating a potentially different effect size when there is a large heterogeneity of results, smaller experiments are still informative about the distribution of effect sizes (as their effect sizes are also feasible). In turn, both small and large experiments tend to be regarded as being more equally informative in random-effects models (even though larger experiments have a slightly greater weight within the joint conclusion \cite{borenstein2011introduction}). 

As the heterogeneity of results is commonplace in SE experiments \cite{juristo2012replication, sjoberg2007future, hayes1999research, miller1999can, hannay2009effectiveness}, many factors may have an impact on SE experiment results \cite{basili1999building}, and experimental changes, opportunistic recruitment of participants and different types of subjects (e.g., professionals vs. students) are typical in groups of SE replications (see Section \ref{background}), we recommend relying by default on random-effects models to provide joint conclusions. 

Specifically, if using AD we suggest the use of random-effects meta-analysis models \cite{borenstein2011introduction}. If using IPD-S, we suggest the use of linear mixed models (LMMs) \cite{brown2014applied}.

\subsubsection{Application to the illustrative group of replications}

\textbf{Activity 3.1. Apply AD.} The application of AD requires calculating the effect sizes---and corresponding variances---of the replications from their summary statistics. They are then pooled using a random-effects meta-analysis model \cite{borenstein2011introduction}. 

\begin{figure}[h!]
    \centering
    \includegraphics[width=9cm,keepaspectratio]{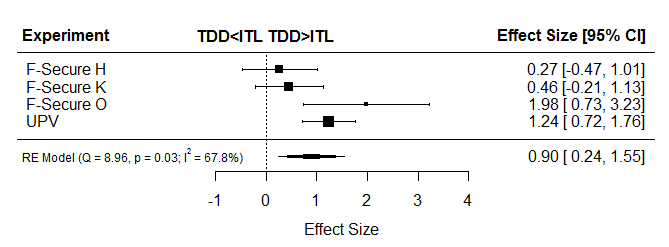}
    \caption{Forest plot: ITL vs. TDD.}
    \label{ad}    
\end{figure}

\textbf{Example.} First, we calculate Cohen's ds---and corresponding variances---of the replications from their summary statistics (i.e., sample sizes, means, standard deviations, and correlations between ITL and TDD \cite{borenstein2011introduction}). As four subjects at UPV have missing data, their data have to be discarded---as they did not provide complete observations for the correlation between ITL and TDD \cite{borenstein2011introduction}. Then, we pool together all Cohen's ds using a random-effects meta-analysis model \cite{borenstein2011introduction}. Figure \ref{ad} shows the forest plot of the meta-analysis. 

As Figure \ref{ad} shows, TDD outperforms ITL in all replications. Additionally, the joint effect size ($M=0.90$) is \textit{large}---according to rules of thumb \cite{borenstein2011introduction}---and statistically significant (as the 95\% CI does not cross 0). Besides, there is, according to rules of thumb, a \textit{medium} ($I^2=67.8\%$) heterogeneity of results \cite{borenstein2011introduction}. Thus, moderators should be identified to explain the detected heterogeneity of results. 

\textbf{Activity 3.2. Apply IPD-S.} IPD-S is straightforward to apply. It is sufficient to fit a LMM with two factors: Treatment and Experiment, considering Treatment as a random effect across the experiments \cite{whitehead2002meta, brown2014applied}.

\textbf{Example.} To analyze the illustrative group of replications with IPD-S, we pool the raw data of all the replications together and then analyze them using a LMM. Table \ref{results_analysis_group} shows the results of the LMM. 

\begin{table}[h!] \centering 
  \caption{LMM results.} 
  \label{results_analysis_group} 
\begin{tabular}{lccc} \hline \hline
\textbf{Factor} & \textbf{Estimate} & \textbf{95\% CI} & \textbf{$p$-value} \\ \hline
ITL & 27.44  & (13.08, 41.79) & $<$0.001\\
TDD & 56.27  & (22.12, 90.42) & $<$0.001\\ \hline \hline
$M_{Diff}$ & 28.83  & (9.72, 47.93) &  0.004 \\ 
$sd_{Diff}$ & 16.09 & & \\ \hline 
\end{tabular} 
\end{table}

As Table \ref{results_analysis_group} shows, the difference in performance between TDD and ITL is relevant ($M_{Diff}=28.83$) and statistically significant ($p$=0.004). Looking at the difference in performance between TDD and ITL (i.e., $M_{Diff}$) and the effect of the control approach (i.e., ITL), we reach the conclusion that \textit{TDD doubles the performance of ITL} (i.e., 56.27/27.44). Unlike AD, participants with missing data have been included to provide joint conclusions \cite{brown2014applied}. Finally, the standard deviation of the differences between TDD and ITL across the replications (i.e., $sd_{Diff}$) is relatively large compared with the overall difference (i.e., $M_{Diff}$): 16.09/28.83=0.56. This suggests that there is heterogeneity. Thus, moderator effects should be identified to explain the observed heterogeneity of results.

\textbf{Summary of example.} AD indicated that there is a large---and statistically significant---joint effect size with medium heterogeneity. Also, AD was able to visualize that TDD outperformed ITL across all the replications. IPD-S showed that TDD doubled the performance of ITL. It also meant that we could include missing data when providing joint conclusions and confirm the statistical significance of results observed with AD.

\subsubsection{Guideline 3: Use AD and IPD-S}

\begin{itemize}

    \item{Use \textit{AD and IPD-S in tandem} to provide joint results. Use AD because of its intuitive visualizations (i.e., forest plots) and straightforward heterogeneity statistics. Use IPD-S because of its ability to convey joint results in natural units, and its flexibility for analyzing replications with missing data.}

    \item{Use \textit{random-effects models} by default to provide joint conclusions. Particularly, use LMMs \cite{brown2014applied} for IPD-S, and random-effects meta-analysis models \cite{borenstein2011introduction} for AD.}
\end{itemize}

What impact may this guideline have on the findings of joint analyses of groups of SE replications? Adherence to this guideline may have potentially resulted in more intuitive joint conclusions for 38\% of the groups of replications (i.e., groups that only applied AD with standardized effect sizes to provide joint conclusions, see Section \ref{limitations}). 

\section{Step 4: Conduct Exploratory Analyses}
\label{step_4}

In Sections \ref{guideline_four}, \ref{guideline_five} and \ref{guideline_six}, we outline the three guidelines that we propose to overcome the most common limitations of groups of SE replications for identifying moderators. The description includes its application to the illustrative group of replications.

\subsection{Use AD plus IPD in tandem to identify experiment-level moderators}
\label{guideline_five}

The identification of experiment-level moderators increases knowledge of software development. We suggest applying both AD and IPD-S in tandem to identify experiment-level moderators. The benefits of using AD plus IPD in tandem have already been discussed in Section \ref{benefits_tandem}. Therefore, we will not repeat our arguments here.

\subsubsection{Application to the illustrative group of replications}

\textbf{Activity 4.1. Identify experiment-level moderators.} To identify experiment-level moderators with AD, perform either a sub-group meta-analysis for \textit{categorical} moderators or a meta-regression for \textit{continuous} moderators. To identify experiment-level moderators with IPD-S, fit LMMs with interaction terms \cite{whitehead2002meta, brown2014applied}.

\begin{figure}[h!]
    \centering
    \includegraphics[width=9cm,keepaspectratio]{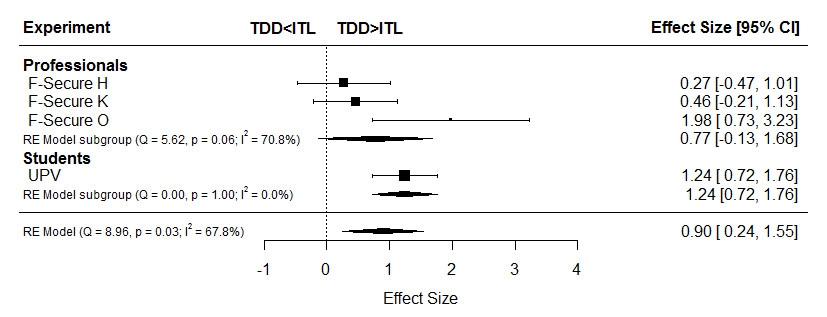}
    \caption{Forest plot: professionals vs. students.}
    \label{sub_group}    
\end{figure}

\textbf{Example.} We run an AD sub-group meta-analysis to assess the effect of the type of subject (i.e., professionals vs. students) on results. Figure \ref{sub_group} shows the forest plot for the sub-group meta-analysis that we performed. Table \ref{results_sub_group} shows the result of the sub-group meta-analysis.

\begin{table}[h!]
\small
\begin{center}
\caption{Sub-group meta-analysis: professionals vs. students.}
\label{results_sub_group}
\begin{tabular}{ l | c | c| c| c} \hline \hline
\textbf{Group} & \textbf{N} & \textbf{Estimate} & \textbf{95\% CI} & \textbf{$I^2$} \\ \hline
Professionals & 3 & 0.77  & (-0.13, 1.68) & 70.8\% \\
Students & 1 & 1.24  & (0.72, 1.76) & 0 \\
Difference & - & 0.47  & (-0.58, 1.52) & - \\ \hline 
\end{tabular}
\end{center}
\end{table}

As Table \ref{results_sub_group} shows, both professionals ($M=0.77$) and students ($M=1.24$) perform better with TDD than with ITL. However, the difference in performance between students and professionals ($M=0.47$) is relevant---\textit{medium} according to rules of thumb \cite{borenstein2011introduction}. In other words, students appear to benefit more from TDD than professionals. However, despite the relevance of the moderator effect, there was a wide 95\% CI. To be precise, the 95\% CI ranges from a \textit{medium} and \textit{negative} effect (i.e., -0.58) to a \textit{large} and \textit{positive} effect (1.52). This results in a non-statistically significant moderator effect due to the small number of replications analyzed. Thus, more replications are needed to increase the precision of experiment-level moderator effects.

Finally, we complement the results of AD with the findings of IPD-S. To do this, we run a LMM with interaction terms \cite{whitehead2002meta, brown2014applied}. Table \ref{results_analysis_moderator_experiment} shows the results of the LMM that we performed.

\begin{table}[!h] \centering 
  \caption{LMM experiment-level moderators: professionals vs. students.} 
  \label{results_analysis_moderator_experiment} 
\begin{tabular}{lccc} \hline \hline
\textbf{Interaction} & \textbf{Estimate} & \textbf{95\% CI} & \textbf{$p$-value} \\ \hline
Type:Students & 16.32  & (-37.16, 69.55) & 0.545 \\ \hline
\end{tabular} 
\end{table}

As Table \ref{results_analysis_moderator_experiment} shows, the difference in performance between students and professionals with TDD appears to be large ($M=16.32$)---at least compared with the difference in performance between TDD and ITL in the main analysis ($M=28.83$). In view of this, students appear to perform around 60\% (i.e. 16.32/28.83) better than professionals using TDD. However, this should be further substantiated with more replications as the 95\% CI ranges from negative to positive results ($95\%~CI=(-37.16, 69.55)$). Again, the group of replications is too small to detect experiment-level moderators.
 
\textbf{Summary of example.} Students appear to benefit more than professionals from TDD. But four replications with 6, 11, 7 and 33 subjects are not enough to detect the effect.

\subsubsection{Guideline 4: Use AD and IPD-S to Identify Experiment-Level Moderators}

Use \textit{AD and IPD-S in tandem} to assess experiment-level moderators.

What benefit may this guideline have on joint analysis practices for groups of SE replications? Forty-two percent of the groups of replications (i.e., groups that adopted a textual approach to eliciting experiment-level moderators, see Section \ref{limitations}) could have achieved more transparent moderator effects in groups of SE replications. Using AD plus IPD in tandem to identify experiment-level moderators should enhance the findings of groups of SE replications.

\subsection{Use IPD to identify participant-level moderators}
\label{guideline_six}

New knowledge is also gained by identifying participant-level moderators. 

\subsubsection{Benefits of using IPD to identify participant-level moderators}

IPD-S is better than AD at identifying participant-level moderators \cite{fisher2011critical, lambert2002comparison}. This is because AD \textit{may be underpowered} if the averaged participant characteristics do not vary much across the replications \cite{lambert2002comparison, debray2015get} and subject to \textit{ecological bias}  when identifying participant-level moderators (i.e., the average effect may not be representative of the effect on the population) \cite{berlin2002individual, fisher2011critical, debray2015get}. This may result in misleading conclusions. Thus, as is already common practice in medicine \cite{fisher2011critical, fisher2017meta}, we recommend relying by default on IPD-S models to identify participant-level moderators.

\subsubsection{Application to the illustrative group of replications}

\textbf{Activity 4.2. Identify participant-level moderators.} To identify participant-level moderators with IPD-S, it is sufficient to fit LMMs with interaction terms \cite{fisher2011critical, fisher2017meta}. As Fisher et al. \cite{fisher2011critical, fisher2017meta} noted, special attention should be paid to separating the variance of moderator effects within and between experiments.

\begin{figure}[h!]
    \centering
    \includegraphics[width=8cm,keepaspectratio]{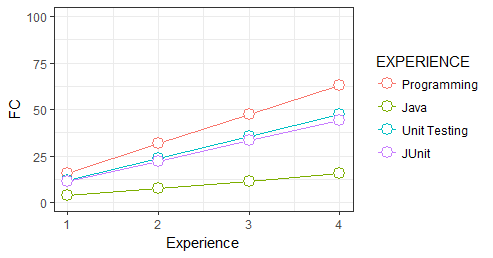}
        \caption{LMM interactions: participant-level moderators.}
    \label{ipd_moderators}
\end{figure}

\textbf{Example.} We ran a series of LMMs with interaction terms to assess the effect of participant experience with programming, Java, unit testing or JUnit on results. Table \ref{results_analysis_moderator_individual} shows the results of the LMMs. Figure \ref{ipd_moderators} shows the regression plot for the moderator effects. 

\begin{table}[!h] \centering 
  \caption{LMM participant-level moderators: participant experience.} 
  \label{results_analysis_moderator_individual} 
\begin{tabular}{lccc}  \hline \hline
\textbf{Interaction} & \textbf{Estimate} & \textbf{95\% CI} & \textbf{$p$-value} \\ \hline
Programming & 15.76 & (0.49, 31.04) & 0.04 \\ 
Java & 3.85 & (-8.13, 15.83) & 0.52 \\
Unit testing & 11.79  & (-5.95, 29.54) & 0.18 \\ 
JUnit & 11.07  & (-6.26, 28.41) & 0.20 \\ \hline 
\end{tabular} 
\end{table}

As Figure \ref{ipd_moderators} shows, the more experienced participants are with programming, Java, unit testing or JUnit, the more TDD outperforms ITL. In other words, TDD appears to perform better for more experienced developers and vice versa. Additionally, as Table \ref{results_analysis_moderator_individual} shows, the significance level for programming experience was lower than 0.1---the significance threshold that we recommended for identifying moderators (see Section \ref{step_4}). Thus, in principle, \textit{programming experience may be moderating the effects of TDD on quality}. However, this information should be regarded with caution because there were inflated Type I error rates (due to the execution of a total of five exploratory analyses) and a wide 95\% CI interval (varying from 0.49, an almost negligible effect, to 31.04, a large effect). Thus, the results may be spurious. Finally, participant experience with programming may also be confounded with other variables (e.g., the age of the developers, their experience with Java, etc.).

\textbf{Summary of example.} Participant experience with programming appears to moderate TDD effects.

\subsubsection{Guideline 5: Use IPD-S to Identify Participant-Level Moderators}

Use \textit{IPD-S to identify participant-level} moderators.

What benefit may this guideline have on the findings of joint analyses of groups of SE replications? Seventy-five percent of the groups of replications (i.e., groups that adopted a textual approach to eliciting participant-level moderators or did not elicit participant-level moderators, see Section \ref{limitations}) could have detected more transparent moderator effects in groups of SE replications. The use of IPD to identify participant-level moderators should enhance the findings of groups of SE replications.

\subsection{Acknowledge limitations of exploratory analyses}
\label{guideline_four}

The limitations of exploratory analyses should be acknowledged.

\subsubsection{Limitations of exploratory analyses}
\label{limitations_exploratory}

Exploratory analyses have three limitations:

    \begin{itemize}

    \item \textit{They are unable to provide cause-effect relationships} because replications are \textit{designed} exclusively to study the effects of the treatments on the response variables. Thus, it is impossible to establish the cause-effect relationships of other variables (e.g., moderators) on the response variables. In SE terms, it is risky to claim that the programming language is the reason for the different results detected using different programming languages across two replications. Note that other variables than the programming language, such as different participant characteristics, for example, may be the real cause for this difference in the results. If any cause-effect claims are to be made about moderators, experiments assessing such questions have to be undertaken beforehand. For instance, a new experiment where participants are randomly assigned to either one or other programming language could serve to identify whether the programming language is the cause of the effects on the results. 

    \item \textit{They increase the risk of committing statistical errors}. Many statistical analyses are typically run to identify moderators (e.g., one per moderator \cite{debray2015get}), which inflates the Type I error rates. In other words, spurious statistical significant results may emerge out of multiple testing merely by chance. Such inflated Type I error rates may need to be corrected with \textit{multiple comparison} correction procedures (such as the Bonferroni correction \cite{quinn2002experimental}). However, this is troublesome in groups of SE replications: on top of their already small sample sizes and the small number of replications (and, thus, poor moderator detectability), even more demanding statistical thresholds are set for identifying moderators. For instance, according to the Bonferroni correction, a statistical threshold of 0.05/3 may be needed to detect moderators and assess three different moderators---one per analysis---. Because of the limitations of groups of SE replications in this regard, we recommend either: (1) setting a statistical threshold of 0.1 to identify moderators---despite the heightened probability of committing statistical errors, or (2) focusing more on the \textit{magnitude} and \textit{sign} of moderator effects---and their corresponding 95\% CIs---rather than on their statistical significance.

    \item \textit{Moderator effects can be confounded if multiple simultaneous changes are made across replications}. For example, if both the programming language and the unit testing tool change simultaneously across two replications, it may be misleading to claim that differences across replication results are solely due to the programming language. The difference in the results may be due to the programming language, the IDE, or a mixture of both. Another typical case of confounding is to claim that differences across replication results are induced by the subject type when different types of subjects are evaluated across two replications (e.g., professionals vs. students) and the replications provide different results. In particular, other variables may also be behind the difference in results such as age (e.g., professionals may be older than students), motivation (e.g., students whose grades are at stake may be more motivated than professionals), treatment conformance (e.g., professionals may deviate from the procedure more than students \cite{dieste2017professionals}), some threats to validity may materialize in some replications and not in others (e.g., drop-outs, missing data, fatigue), etc.

\end{itemize}

Researchers running joint analyses of groups of replications should acknowledge the limitations of exploratory analyses in their papers.

\subsubsection{Application to the illustrative group of replications}

\textbf{Activity 4.3. Acknowledge limitations of exploratory analyses.} To acknowledge the limitations of exploratory analyses, it suffices to check through and adapt the list of limitations that we outlined in Section \ref{limitations_exploratory} to the group of replications---and moderators---that are to be investigated.

\textbf{Example.} We plan to investigate the effect of \textit{one} experiment-level moderator (i.e., type of subject, students vs. professionals) and \textit{four} participant-level moderators (i.e., experience with programming, Java, unit testing, and JUnit) on results. We acknowledge that none of the above moderators may be the real reason behind the detected heterogeneity of results, and other confounding variables may also be responsible for the detected difference in results. For instance, students appeared to be more motivated than professionals, students adhered more closely to the TDD process, professionals were older than students, etc. Also, we acknowledge that the chance of achieving spurious statistically significant results increases because we intend to run five data analyses (i.e., one per moderator, following Fisher's approach \cite{fisher2011critical}) \cite{quinn2002experimental}. This may invalidate the conclusions reached. Thus, we will only use exploratory analyses as a way of motivating further research and never to draw definite conclusions \cite{lau1998summing}.

\textbf{Summary of example.} The group of replications is too small for making definite claims about the results regarding subject type. Additionally, results for programming experience may be spurious due to the wide 95\% CI that materialized. Finally, programming experience may be confounded with other variables (e.g., age of the participants, experience with Java, etc.).

\subsubsection{Guideline 6: Acknowledge Limitations of Exploratory Analyses}

Separate exploratory analyses from the main analysis and acknowledge their limitations.

What benefit may this guideline have on the findings of joint analyses of groups of SE replications? Thirty-eight percent of the groups of replications that did not perform exploratory analyses (see Section \ref{limitations}), and 88\% of the groups of replications that did not acknowledge the limitations of exploratory analyses could have identified informative moderator effects in groups of SE replications. The performance of exploratory analyses and acknowledgment of their limitations should enhance the findings of groups of SE replications.

\section{Threats to Validity}
\label{threats}

\textit{We focused on the aggregation of quantitative results. What about qualitative results (e.g., text transcripts, etc.)?} Throughout this article we focused exclusively on the aggregation of quantitative results into joint conclusions. We acknowledge that this limits the applicability of our guidelines. However, we decided to focus on quantitative results because SE experiments are usually coupled with the acquisition and analysis of this type of results \cite{wohlin2012experimentation, kitchenham2015evidence, stol2018abc} and most of the groups of replications uncovered by our SMS only aggregated quantitative results \cite{adrisms}. For an overview of the methods that can be used for aggregating qualitative ---or qualitative and quantitative results---into joint conclusions, we refer interested readers to Cruzes and Dyba \cite{cruzes2011research} and Kitchenham et al. \cite{kitchenham2015evidence}. They discuss meta-ethnography, narrative synthesis, qualitative cross-case analysis, thematic analysis, meta-summary, vote counting, grounded theory, content analysis, case survey, qualitative comparison analysis, aggregated synthesis, realist synthesis, meta-synthesis and meta-study. The aggregation of qualitative and quantitative results is discussed in \cite{popay2006guidance, dixon-woods2001qualitative, thomas2004integrating}.

\textit{We used only one analysis procedure. Are there any limitations to the way in which we tailored it to SE?} Like \cite{whitehead2002meta, borenstein2011introduction}, we acknowledge that it is unfeasible to provide definite guidance on  aggregation techniques and statistical models for use across the board---independently of the characteristics of the data or the intentions of the analyst (e.g., to provide joint results or to identify moderators)---. However, we tried our best to provide tailored statistical advice for aggregating the results of groups of SE replications considering their common characteristics and the limitations on joint data analysis based on the guidelines typically followed in medicine and pharmacology to analyze MCTs \cite{bero1995cochrane, anello2005multicentre, lewis1999statistical, stewart2015preferred} and our understanding of the statistical methods that can be used in circumstances that are typical of groups of SE replications---at least according to well-known references in medicine, pharmacology, and the social sciences \cite{debray2015get, whitehead2002meta, maas2005sufficient, mcneish2016effect, bell2015explaining}. We acknowledge that this is only a first approximation towards tailoring a definite analysis procedure and that further research is needed in order to provide more evidence on the suitability of the analysis procedure for analyzing groups of SE replications.

\textit{We gathered the references on data analysis opportunistically. Might not this introduce bias?} Unfortunately, we could not systematically gather references on the topic of how to analyze MCTs---or groups of replications with characteristics typical in SE---: the number of articles retrieved from online databases with the terms "aggregation" and "experiments" was unmanageable. Thus, we acknowledge that our guidelines may be open to bias. However, we made every effort to consult reliable resources on the topic of aggregation of experiment results from both mature experimental disciplines, such as medicine and pharmacology, and other areas, such as social research, education and econometrics. We also strove to embed guidelines providing the results not only of single aggregation techniques but also of different aggregation techniques applied in tandem (see the recommendation to use IPD-S and AD in tandem). This should reduce the potential bias that may have been introduced due to the non-systematic selection of references on data analysis.  

\textit{We selected random-effects models over fixed-effects models. Are there not any limitations to this advice?} Contrary to medicine, where fixed-effects models are encouraged for use by default \cite{whitehead2002meta, anello2005multicentre, phillips2003e9}, we recommend the use of random-effects models instead \cite{borenstein2011introduction, whitehead2002meta, greco2013meta}. This is a controversial recommendation: some authors from other disciplines suggest that a minimum of five \cite{feaster2011modeling}, ten \cite{snijders2011multilevel, mcneish2016effect}, fifteen or even more experiments \cite{mcneish2016effect, maas2005sufficient, duncan1998context} are needed to obtain reliable \textit{variance} parameter estimates in random-effects models. However, we recommend the use of random-effects models by default as SE is commonly concerned with providing joint results (i.e., differences between \textit{means}) rather than making inferences on variance parameters. Additionally, random-effects models tend to provide more conservative results than fixed-effects models (i.e., 95\% CIs tend to be wider with random-effects models \cite{whitehead2002meta, petitti2000meta, chen2013applied}), and random-effects models produce identical results to fixed-effects models when there is no heterogeneity \cite{borenstein2011introduction}. Finally, sensitivity analyses assessing the robustness of results to the specification of fixed-effects models---rather than random-effects models---can also be run in groups of SE replications \cite{borenstein2011introduction}. We refer interested readers to Thabane et al. \cite{thabane2013tutorial}.  

\textit{We did not check the statistical assumptions of the statistical tests used ($t$-tests, LMMs). Is this not a limitation?} As usual after analyzing the data of individual experiments \cite{wohlin2012experimentation}, it is also necessary to check the statistical assumptions of the statistical tests used after aggregating the results \cite{brown2014applied, hox2010multilevel}. For example, if LMMs are fitted to analyze the data, then the normality assumption needs to be checked \cite{whitehead2002meta, hox2010multilevel}. We acknowledge that SE data may be not normal and concede that there are more advanced statistical methods for analyzing non-normal data (see Section \ref{alternative}). However, we resorted to $t$-tests and LMMs in this article because they are robust to departures from normality \cite{fagerland2012t, mcculloch2011misspecifying}, especially with larger sample sizes---as is the case when the raw data of all the replications are pooled together---\cite{lumley2002importance}. In any case, as model diagnostics procedures are standard across the disciplines, we refer interested readers to specialized literature on the topic \cite{brown2014applied, hox2010multilevel, whitehead2002meta, mcculloch2001generalized}. 

\textit{We use only one illustrative group of replications. Is the analysis procedure applicable in other cases? } For reasons of space, we illustrated the application of the analysis procedure to only one group of replications. We admit that this places some limitations on the generalizability of the analysis procedure. However, we tried to select what is, according to the results of a previous SMS addressing this issue, a representative group of SE replications \cite{adrisms}. Accordingly, we expect our guidelines to be applicable to the analysis of a large percentage of groups of SE replications. Additionally, we also provide further references in Section \ref{alternative} indicating how to analyze groups of replications with different experimental designs and data characteristics.

\section{Alternative Experimental Designs}
\label{alternative}

As typical in groups of SE replications, this article analyzes a group of replications where all the replications have an identical experimental design: an AB \textit{within-subjects} design (i.e., a design where the participants apply first Treatment A and then Treatment B in a later session \cite{wohlin2012experimentation}). It is straightforward, based on our procedure, to analyze groups of replications where all the experiments have an identical AB \textit{between-subjects} design (i.e., a design where the participants are randomly assigned to either Treatment A or B): (1) for IPD-S, it is sufficient to remove the repeated-measures structure at participant level \cite{whitehead2002meta}; (2) for AD, it is sufficient to adapt the effect size variance formulae to the between-subjects design \cite{borenstein2011introduction}. Besides, as groups of replications with AB between-subjects designs are commonplace in medicine, there is no shortage of references indicating how to analyze such designs with both IPD-S \cite{debray2015get, whitehead2002meta, fisher2011critical} and AD \cite{borenstein2011introduction}.

Whenever groups of replications contain a mixture of AB between-subjects designs and AB within-subjects designs, researchers can still use LMMs with the IPD-S approach to provide joint conclusions---as LMMs can account for replications with missing data, provided that they are missing at random (e.g., when it is possible to tell from the experimental design whether the subject data will or will not be missing in a specified experimental session \cite{kwok2008analyzing, hoffman2007multilevel, maas2003multilevel}). AD can also be used for analyzing mixtures of AB between-subjects designs and AB within-subjects designs \cite{morris2002combining}: it is sufficient to calculate a consistent pooled standard deviation for standardizing Cohen's d (typically the average standard deviations of Treatments A and B) and then select the appropriate variance formulae depending upon the experimental design (i.e., a within-subjects or between-subjects design \cite{morris2002combining}). As experiments with different experimental designs may be estimating different true effect sizes \cite{morris2002combining}, however, exploratory analyses investigating the difference of results across experimental designs \cite{morris2002combining, thabane2013tutorial} (e.g., by means of a sub-group meta-analysis \cite{borenstein2011introduction}) should be used.

Throughout this article, we analyzed experiments whose response variable was measured on a \textit{continuous scale}, which is typical in SE \cite{adrisms}. Also, we relied on the statistical tests that we ran being robust to departures from normality \cite{fagerland2012t,mcculloch2011misspecifying}. Still, researchers may question the reliability of their inferences if data largely depart from the normality assumption \cite{kitchenham2016robust}. If this is the case, researchers may resort to data transformation (e.g., Box-Cox transformations) to make the normality assumption hold and then apply IPD-S or AD to provide joint conclusions \cite{quinn2002experimental}. Researchers may also resort to more advanced statistical techniques such as bootstrapping to provide inferences with both IPD-S \cite{ren2010nonparametric, field2007bootstrapping} and AD \cite{nakagawa2007effect}. Finally, researchers applying AD can also use non-parametric effect sizes such as Cliff's delta to provide joint conclusions \cite{kitchenham2016robust}. Eventually, if data do not meet the homogeneity of variances assumption, researchers can opt for generalized least squares (GLS) models under the IPD-S umbrella (because they can accommodate different variance terms per treatment \cite{zuur2009mixed}) or  Glass's delta---rather than Cohen's d---under the AD umbrella \cite{pautz2018use}.

Also, response variables may be measured on a non-continuous scale (e.g., binary, count, etc. \cite{quinn2002experimental}). In this case, generalized linear mixed models (GLMMs) may be more appropriate than LMMs for analyzing the data with IPD-S \cite{quinn2002experimental}. We refer interested readers to Zuur et al. \cite{zuur2009mixed, zuur2013beginner} for an accessible introduction to the topic using the R programming language. Pautz et al. \cite{pautz2018usenon} and Fritz et al. \cite{fritz2012effect} provide illustrative examples of how to calculate effect sizes---to be later combined with AD---for non-continuous response variables. 

In multilevel data structures typical of groups of SE replications---where subjects are nested within replications and subjects can be measured several times (once or more per treatment) throughout the experiment---, data are typically correlated within \textit{clusters} (i.e., clusters of both replications and participants \cite{whitehead2002meta}). This correlation needs to be taken into account when analyzing the data (i.e., by including the clustering units as either fixed factors or as random factors with the selection of appropriate variance-covariance matrices \cite{whitehead2002meta}). Throughout this article, we relied on random factors---for both participants and replications---and the variance-covariance matrix used by the \textit{lme} R function when fitting LMMs: the \textit{unstructured} variance-covariance matrix \cite{finch2016multilevel}. With this variance-covariance matrix, both the treatment and control (e.g., ITL in the illustrative group of replications) effects are assumed to be correlated across the replications \cite{finch2016multilevel}. If this does not hold, however, other variance-covariance matrices may be more suitable for analyzing the group of replications (e.g., by assuming independent treatment and control approach effects). For an accessible introduction to the topic, we refer interested readers to Finch et al. \cite{finch2016multilevel} and Zuur et al. \cite{zuur2009mixed}.

Finally, missing data can materialize in SE experiments due to drop-outs or protocol deviators. Under these circumstances, researchers may resort to LMMs under the IPD-S umbrella to aggregate results (LMMs can be used to analyze groups of replications with missing data as long as data are missing at random \cite{twisk2013multiple}). They may also rely on imputation methods (e.g., single imputation, multiple imputation, etc. \cite{schafer2002missing}) to analyze the data with other IPD-S models that do not permit the inclusion of missing data (e.g., repeated-measures ANOVA) or AD. Whichever procedure is finally selected for handling missing data, sensitivity analyses should be conducted to ensure that the results are robust to the missing data procedure specification \cite{thabane2013tutorial}. We refer interested readers to Little et al. \cite{little2012prevention} and Schafer and Graham \cite{schafer2002missing}.

\section{Related Work}
\label{related_work}

To the best of our knowledge, no previous attempts have been made in SE to provide guidelines for analyzing groups of SE replications---when researchers have access to the raw data and first-hand knowledge of the settings and participant characteristics. However, some previous articles already discussed the suitability of various research synthesis methods for combining \textit{published results}. For instance, in the late 1990s, Pickard et al. \cite{pickard1998combining} outlined the advantages and disadvantages of meta-analysis of effect sizes \cite{borenstein2011introduction}, Fisher's method (i.e., an aggregation of $p$-values technique \cite{borenstein2011introduction}) and vote-counting (i.e., a form of narrative synthesis procedure \cite{borenstein2011introduction}) for aggregating the results of a series of case studies. However, Pickard et al. \cite{pickard1998combining} aggregated the results of case studies whose raw data were accessible---because they argued that the effect size that they used (i.e., the Pearson correlation coefficient \cite{borenstein2011introduction}) needed to be computed from the raw data to guarantee the consistency of results across the studies \cite{pickard1998combining}. Ultimately, Pickard et al. also acknowledged that meta-analysis could be performed as long as study reports provided appropriate summary statistics to backcalculate the necessary effect sizes \cite{pickard1998combining}.

Ever since, meta-analysis has been tightly coupled in SE with the concept of synthesizing the results of already published studies (i.e., typically with standardized effect sizes such as Cohen's d and AD) \cite{hayes1999research, miller2000applying, fernandez2007aggregation, shepperd2018role, kitchenham2015evidence}. To do this, researchers should backcalculate appropriate effect sizes from study reports, and, if the studies are not very dissimilar, use either fixed-effects models or random-effects models for combination \cite{fernandez2007aggregation, hayes1999research, kitchenham2015evidence, sjoberg2007future}. However, disparate advice with regard to the use of meta-analysis can be found in the SE literature. For example, while Pickard et al. \cite{pickard1998combining} acknowledge that meta-analysis is only appropriate when the studies are homogeneous enough---or when the heterogeneity across the studies can be clearly attributed to certain conditions \cite{pickard1998combining}---, Miller et al. \cite{miller2000applying} indicate that identical studies (e.g., replications using identical materials) may result in "strong correlations" affecting the reliability of the joint conclusions. This view has also been backed up by others in the SE community \cite{kitchenham2008role}. At the same time, and given the commonly heterogeneous results reported in the literature and the myriad variables that typically change across SE studies, Miller et al. \cite{miller2000applying} finally conclude that \textit{"...the heterogeneity of current empirical results is a major limitation to our ability to apply meta-analytic procedures..."}. 

Due to the limitations of meta-analysis and the particularities of SE studies, other SE researchers have proposed the use of other aggregation techniques for synthesizing already published empirical study results \cite{fernandez2007aggregation, kitchenham2015evidence, olorisade2013determining}. Briefly, such techniques commonly involve some sort of vote-counting technique (e.g., counting positive vs. negative results, small vs. large results, etc.), or the application of different aggregation techniques (e.g., meta-analysis, vote-counting, etc.) depending upon the characteristics of the studies being aggregated (number of available studies, number of changes made across the studies, etc. \cite{fernandez2007aggregation}).

Similar concerns about the limitations of meta-analysis have been also raised in other disciplines over the years \cite{lau1998summing, ioannidis2008research, gurevitch2018meta}. The overall consensus nowadays seems to be that meta-analysis of effect sizes (i.e., AD) should be preferred over narrative synthesis or vote counting techniques---at least for aggregating quantitative results \cite{cooper2009relative, borenstein2011introduction, biondi2016umbrella, gurevitch2018meta}. Also, the meta-analysis of raw data (i.e., IPD-S) outperforms AD in some circumstances (e.g., in terms of statistical flexibility or for identifying participant-level moderators \cite{bero1995cochrane, stewart2002ipd, fisher2017meta}). Still, the debate about the limitations of meta-analysis techniques and research synthesis is ongoing \cite{gurevitch2018meta}.

\section{Conclusions}
\label{conclusions}

Researchers from different groups and institutions are collaborating on the construction of groups of replications in SE. Applying unsuitable aggregation techniques to analyze groups of SE replications may undermine their potential to provide in-depth insights from experiment results.

We learned about the recommendations and guidelines used to analyze and report groups of replications in mature experimental disciplines such as medicine and pharmacology \cite{bero1995cochrane, anello2005multicentre, lewis1999statistical}. Unfortunately, such guidelines could not be directly imported for the analysis of groups of SE replications because of the noticeable differences between groups of replications in SE and medicine that we came across (i.e., in terms of the number of changes made across the replications, participant heterogeneity, statistical power, etc.). 

We designed an analysis procedure with a set of embedded guidelines to analyze the stereotypical group of SE replications \cite{adrisms}. To do this, we adopted the same basic structure typically followed in medicine and pharmacology for analyzing groups of replications. However, we adapted the steps to the characteristics of groups of SE replications, and their common limitations with regard to joint data analysis. The analysis procedure that we propose outlines a minimum set of steps that may potentially increase the informativeness of joint conclusions and moderator effects. It all boils down to providing appropriate descriptive statistics and visualizations to ease the interpretation and incorporation of results into prospective studies, as well as taking advantage of the raw data to provide joint conclusions and identify moderators. AD and IPD-S---random-effects models---are crucial for this purpose. Table~\ref{summary_techniques} shows a summary of how to use the aggregation techniques proposed in our procedure.

\begin{table*}[t!] \centering 
  \caption{Summary of aggregation techniques to be used.} 
  \label{summary_techniques} 
\begin{tabular}{lll} \hline \hline 
 & \textbf{AD} & \textbf{IPD-S} \\ \hline 
\multirow{3}{3cm}{Recommended for} & - Results aggregation & - Results aggregation \\ 
& - Experiment-level moderators & - Experiment-level moderators \\ 
&  & - Participant-level moderators \\ \hline
\multirow{4}{3cm}{When can be used} & - RV metrics can be different   & - RV metrics are identical \\ 
& - Complete observations  & - Allows missing data\\ 
& -	Effect sizes (or raw data to calculate them) & - Raw data are available\\
& are available & \\ \hline
\multirow{2}{3cm}{How should be used} & \multirow{2}{*}{- Fit random effects models} & -	Use Linear Mixed Models (LMMs) \\
& & - Fit random effects models \\ \hline
\multirow{5}{3cm}{How should be used for moderator analyses} & Use either: & -	Fit LMMs with interaction terms \\
& \multirow{3}{6.5cm}{- Sub-group meta-analysis for categorical moderators} & - Increase statistical significance threshold to 0.1 \\ 
& & \multirow{2}{6.5cm}{- Pay less attention to p-values and focus on effect sizes and 95\% CIs}\\
& \multirow{2}{6.5cm}{- Meta-regression for continuous moderators} &  \\ 
& & - Run one analysis per moderator \\ \hline \\
\end{tabular} 
\end{table*}

To wrap up, we encourage SE researchers analyzing groups of replications to justify their aggregation techniques and, more importantly, to transparently report the statistical models and the raw data that they used to provide joint conclusions and identify moderators. With the aim of easing the application of the analysis procedure and with a view to reproducibility, the supplementary material of this article includes the step-by-step commented R code and raw data---with the associated R notebook---that led to the results reported throughout this article. In addition, we offer a more technical tutorial including R code snippets, dataset descriptions, and mathematical formulae to complement the understanding of the R code that we provide. All the supplementary material is also available at figshare (URL: {\color{blue} \url{https://doi.org/10.6084/m9.figshare.7583909.v7}}). We hope this encourages others to give the analysis procedure a go. 

\section*{Acknowledgments}

This work was partially funded by Spanish Ministry of Science, Innovation and Universities research grant PGC2018-097265-B-I00.

\section*{Supplementary Material}

\noindent\fbox{
  \parbox{8.5cm}{
    \textbf{Supplementary material 1:} Raw data (XLSX 13 kb).
    
    \textbf{Supplementary material 2:} Characteristics (XLSX 9 kb).
    
    \textbf{Supplementary material 3:} R code (R 17 kb).
    
    \textbf{Supplementary material 4:} R notebook (Rmd 20 kb).
    
    \textbf{Supplementary material 5:} R Tutorial (PDF 316 kb).
    
  }
}

\ifCLASSOPTIONcaptionsoff
  \newpage
\fi

\bibliographystyle{IEEEtran}
\bibliography{biblio}

\begin{IEEEbiography}
[{\includegraphics[width=1in,height=1.25in,clip,keepaspectratio]{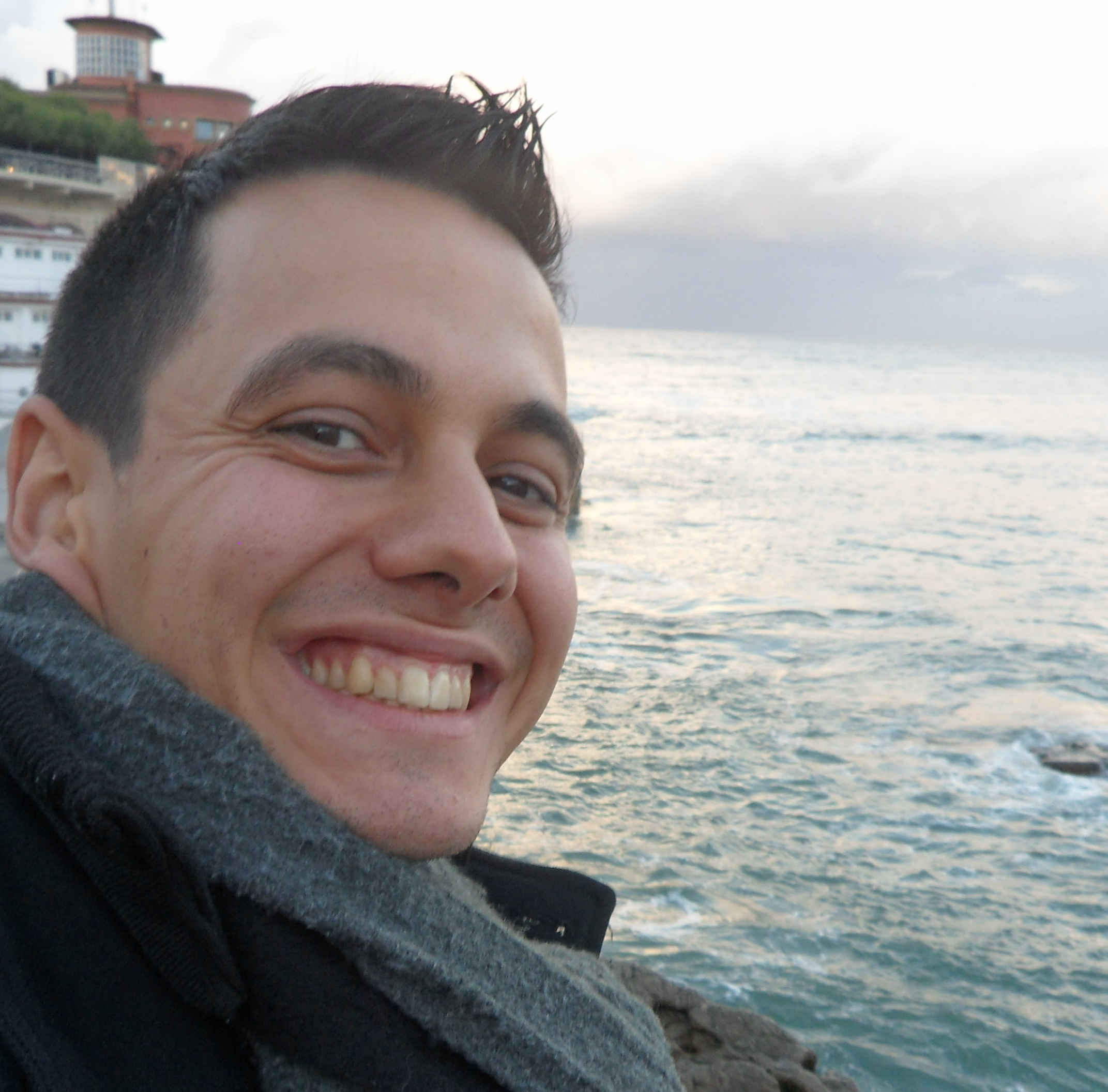}}]{Adrian Santos}
received his MSc in Software and Systems and MSc in Software Project Management from the Technical University of Madrid, Spain, and his MSc in IT Auditing, Security and Government from the Autonomous University of Madrid, Spain. He is a PhD student at the University of Oulu, Finland. He is a member of the American Statistical Association (ASA) and the International Society for Bayesian Analysis (ISBA). 
\end{IEEEbiography}
\begin{IEEEbiography}
[{\includegraphics[width=1in,height=1.25in,clip,keepaspectratio]{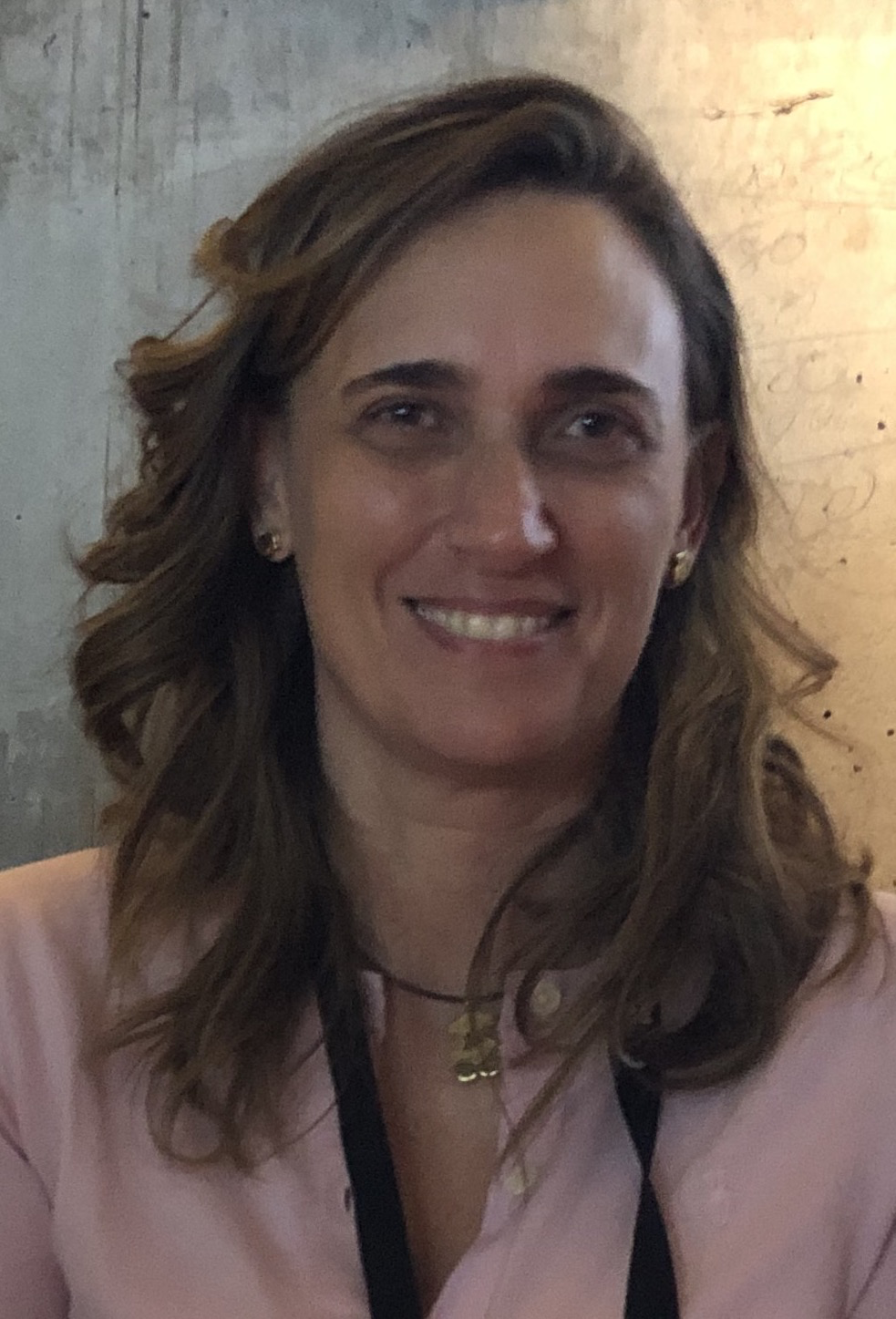}}]{Sira Vegas} has been associate professor of software engineering with the  School of Computer Engineering at the Technical University of Madrid, Spain, since 2008. Sira belongs to the review board of IEEE Transactions on Software Engineering, and is a regular reviewer of the Empirical Software Engineering Journal. She was program chair for the International Symposium on Empirical Software Engineering and Measurement in 2007.
\end{IEEEbiography}
\begin{IEEEbiography}
[{\includegraphics[width=1in,height=1.25in,clip,keepaspectratio]{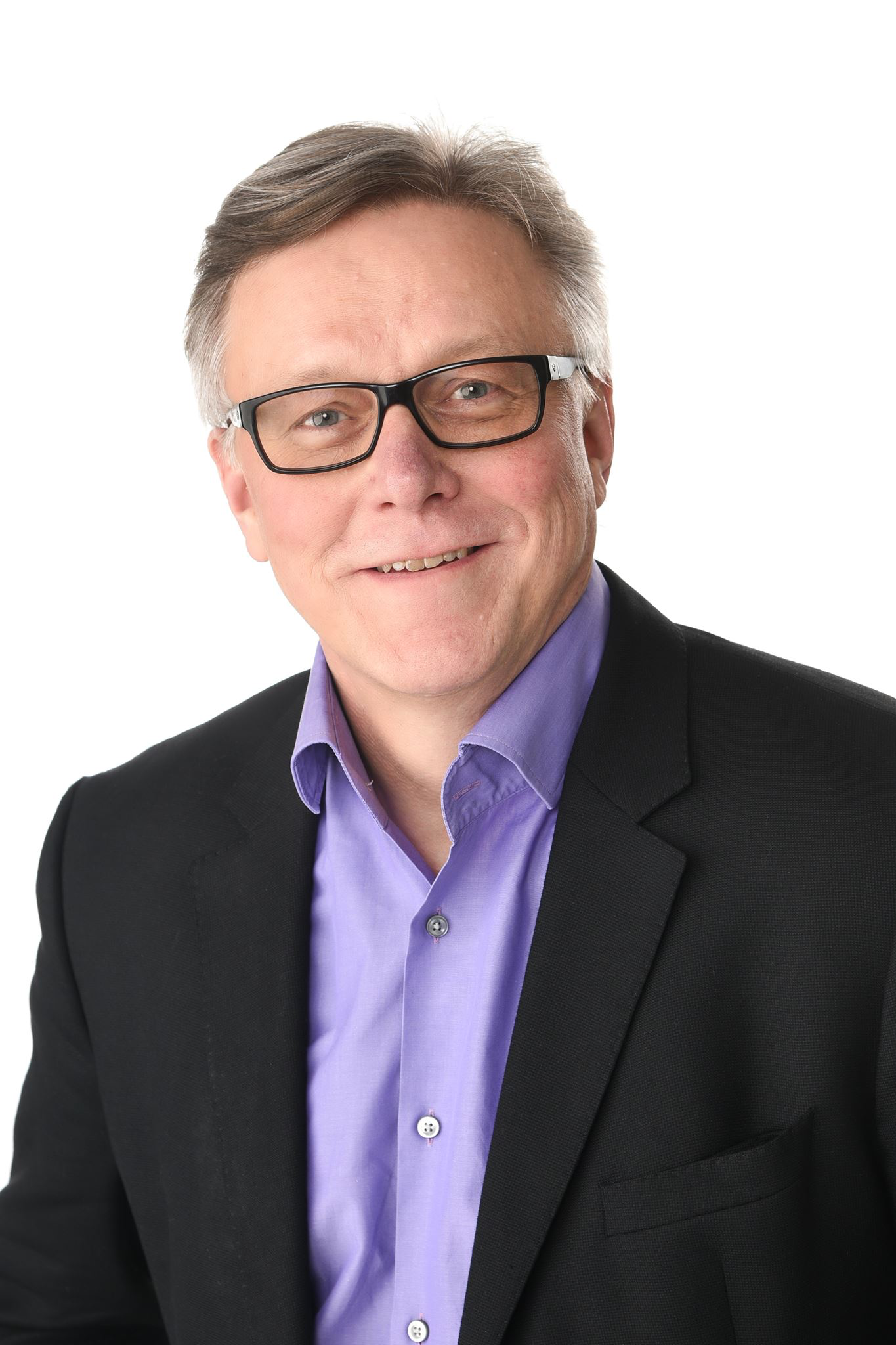}}]{Markku Oivo}
is professor and head of the M3S research unit at the University of Oulu, Finland. He held visiting positions at the University of Maryland (1990-91), Schlumberger Ltd. (Paris 1994-95), Fraunhofer IESE (1999-2000), University of Bolzano (2014-15), and Universidad Polit\'ecnica de Madrid (2015). 
\end{IEEEbiography}
\begin{IEEEbiography}[{\includegraphics[width=1in,height=1.25in,clip,keepaspectratio]{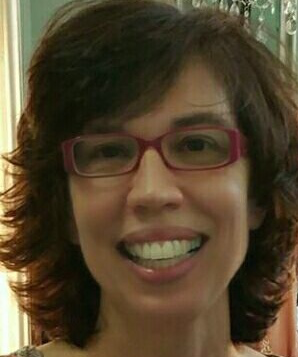}}]{Natalia Juristo} has been full professor of software engineering with the  School of Computer Engineering at the Technical University of Madrid, Spain, since 1997. She was awarded a FiDiPro (Finland Distinguished Professor Program) professorship at the University of Oulu, from January 2013 until June 2018. Natalia belongs to the editorial board of EMSE and STVR. In 2009, Natalia was awarded an honorary doctorate by Blekinge Institute of Technology in Sweden.
\end{IEEEbiography}
\vfill




\end{document}